\documentclass[12pt]{article}
\usepackage{amsmath, amssymb, amsthm, graphicx, booktabs, hyperref, rotating, enumitem, adjustbox,bm} 
\usepackage{array,stmaryrd}
\usepackage{mathtools}
\usepackage{tabularx}
\usepackage{pdflscape}
\usepackage{adjustbox}
\usepackage{subfig}
\usepackage{longtable}
\usepackage{listings}
\usepackage{multirow}
\usepackage{xr}
\usepackage{blkarray}
\usepackage{xcolor}
\usepackage{comment}
\usepackage{natbib}
\usepackage{algorithm}
\usepackage{algpseudocode}
\usepackage[font=small,labelfont=bf]{caption}
\usepackage{authblk}

\newtheorem{theorem}{Theorem}

\newcommand{\blind}{0}

\begin{document}

\def\spacingset#1{\renewcommand{\baselinestretch}%
{#1}\small\normalsize} \spacingset{1}

\if0\blind
{
  \title{\bf PCA score regression: the art of losing power}
  \author[1]{Yu Lu\thanks{Email: ylu136@jhu.edu}}
  \author[2]{Nidhi Pai}
  \author[2]{Erjia Cui}
  \author[1]{Ciprian Crainiceanu}
  
  \affil[1]{Department of Biostatistics, Johns Hopkins University, Maryland, USA}
  \affil[2]{Division of Biostatistics and Health Data Science, University of Minnesota, Minnesota, USA}
  
  \maketitle
} \fi

\if1\blind
{
  \bigskip
  \bigskip
  \bigskip
  \begin{center}
    {\LARGE\bf PCA score regression: the art of losing power}
  \end{center}
  \medskip
} \fi

\bigskip
\begin{abstract}
The regression of principal component scores (RPCS) on covariates is a widely used analytic approach to detect and test for associations between functional measurements and study participant characteristics. Here we show that: (1) RPCS loses power relative to Function on Scalar Regression (FoSR); (2) the amount of power loss depends on the correlation between the PCs and the true effect; (3) if not corrected for multiplicity, RPCS has inflated $\alpha$-level; and (4) current RPCS methods do not provide valid inference for the true effect. In contrast, we show that Function on Scalar Regression (FoSR) can avoid these problems using a particular combination of modeling tools. We validate these theoretical findings through extensive simulations and illustrate their practical implications using minute-level accelerometry data from the National Health and Nutrition Examination Survey (NHANES).
\end{abstract}

\noindent%
\vfill

\spacingset{1.75} 
\section{Introduction}\label{sec:intro}
Understanding the relationship between functional data and scalar covariates is a central challenge in many scientific fields, including the analysis of brain connectivity \citep{reissfosr}, diffusion brain imaging \citep{goldsmith2012longitudinal,greven2010longitudinal,staicu2012}, seismic ground motion \citep{bauer2018}, CD4 counts in studies of HIV infection \cite{fan2000}, reproductive behavior \citep{chiou2004,chiou2003}, carcinogenesis experiments \citep{morris2006}, knee kinematics \citep{antoniadis2007}, human vision \citep{ogden2010}, circadian analysis of cortisol levels \citep{guo2002}, mass spectrometry proteomic data \citep{morris2008,zhumorris2011,zhumorris2012,morris2012}, eye scleral displacement induced by intraocular pressure \citep{leemorris2019}, feeding behavior of pigs \citep{gertheiss2015pigs}, phonetic analysis \citep{aston2010}, and continuous glucose monitoring \citep{gaynanova2020,sergazinov2023}. Analyzing such data requires methods that account for their inherent smoothness, high dimensionality, and correlations over the domain of observation. 

Studying the association between a functional outcome and scalar predictors (e.g., age, gender, BMI) can be conducted using regression of subject-specific principal component scores (RPCS) on covariates \citep{crainiceanu2024book,kokoskareim2017,ramsaysilv2005,ramsayhook2009}. This two-step procedure has been extensively used in practice to study conditions that can influence industrial product performance \citep{McMullen2024, McMullenJurica2024}, associations between complex movement patterns and weightlifting performance \citep{Kipp2012}, coordination differences as a function of rowing expertise \citep{Warmenhoven2017}, and the effect of running mechanics on performance \citep{coffey2011}. In the biomedical and behavioral sciences, it has been widely applied to evaluate associations between daily step count trajectories and clinical outcomes \citep{kringle2023}, phenotype apathy in Alzheimer's disease \citep{zeitzer2013}, and predict changes in sleep, cognition, and mortality from daily accelerometer patterns \citep{zeitzer2018}. Beyond the scope of functional data, this two-step ``PCA-then-regression" is fundamental in modern genetics. In population genetics, eigenvector scores capturing ancestry are regressed on individual SNPs or geographic coordinates to identify loci under selection \citep{chen2016eigengwas, Sharma2022, novembre2008}, while in systems genomics, ``module eigengenes" are regressed on clinical covariates to link gene co-expression patterns to disease states \citep{Langfelder2007, Marchi984, Sabik2020, Tan2013}. \citep{Aschard2014} compared the power of various PCA-based strategies and stated testing only the top PCs often has low power, whereas combining signal across all PCs can have greater power under human genetics context. Recent advances in single-cell transcriptomics have further extended this by using latent factor models to regress cell-specific scores on experimental covariates \citep{AE2025}. These are just a few example of the exploding number of papers featuring PCA followed by regression of the PC scores. 

An alternative to RPCS is to conduct Function on Scalar Regression (FoSR); for a review see Chapter 5 in \cite{crainiceanu2024book}. FoSR was first popularized by \cite{ramsaysilv2005} who introduced it as a functional linear model with a functional response and scalar covariates. The FoSR nomenclature used in this paper was introduced by \cite{reissfosr,reiss2010}. There are many different approaches for conducting inference in this context \citep{cui2022fui,goldsmith2011,scheipl2015famm}, but the general idea is to write down an explicit model for the conditional mean of the functional outcome given the covariates. Many approaches are focused on estimation, while some provide inference based on explicit modeling of the functional residuals \citep{goldsmith2011,scheipl2015famm} or by bootstrapping the subjects \citep{cui2022fui}. 

While recent (yet unpublished) work has demonstrated the advantages of FoSR over RPCS when evaluating time-varying intervention effects in physical activity studies \citep{pai2026}. It is currently unknown exactly what the differences between RPCS and FoSR, what are the theoretical underpinnings of these differences, and what are the practical consequences of these differences. This paper is focused on: (1) providing the modeling and theoretical infrastructure for comparing the two approaches; (2) explaining the relationship between the two estimands and implied hypotheses; (3) providing insights into why and when RPCS loses power; (4) show under what conditions RPCS has no power; and (5) show that these theoretical findings have direct and consequential practical implications in real applications. We will focus on comparing RPCS with FoSR based on penalized spline regression and mixed effects modeling.

The rest of the paper is organized as follows. Section \ref{sec:method} compares the two methos from a theoretical perspective. Section \ref{sec:simulations} presents two simulation studies comparing the power and size of thes two methods. We then compared these two methods on the NHANES study and conclude with a discussion in Section \ref{sec:discussion}.

\section{Methods}\label{sec:method}
The observed data for participant $i$ is $[\{W_i(s), s\in S\}, \mathbf{X}_i,i = 1,\ldots, N]$, where $S=\{s_1,\ldots,s_P\}$ represents a set of observation locations, $W_i(s_p)$ are functional observations for subject $i$ at location $s_p\in S$, and $\mathbf{X}_i = [X_{i1}, X_{i2}, \ldots, X_{iQ}]^t$ is a $Q\times 1$ dimensional vector of scalar covariates. Note that the locations in $S$ are not required to be equally spaced. Let $\mathbf{W}$ be the $N\times P$ data matrix where the $(i,j)$ entry corresponds to $W_{i}(s_j)$. Our primary objective is to quantify the association between the functional response $W_i(s)$ and the scalar covariates $\mathbf{X}_i$. A common analytic pipeline involves applying Principal Component Analysis (PCA) or Singular Value Decomposition (SVD) to the matrix $\mathbf{W}$ to extract subject-specific scores, which are then regressed on $\mathbf{X}_i$. This approach initially ignores the dependence of scores on covariates when conducting PCA (or SVD) and subsequently explicitly models the association between scores and covariates. This procedure is easy to understand and implement, which likely accounts for its popularity. However, we will show that it is equivalent to fitting a 
misspecified model that leads to information and power loss.

\subsection{Function on scalar regression}\label{subsec:FoSR}
To illustrate that, we introduce a rigorous statistical modeling framework designed to frame the question in terms of observable data. This will allow us to show: (1) that the RPCS is always less powerful than direct modeling of fixed effects; (2) when and why this is the case; (3) that the power to detect signals could be zero even when the true effects are large; and (4) how to identify when such power loss has likely occurred. Indeed, consider the case when the true data generating mechanism is the Function on Scalar Regression (FoSR) \citep{crainiceanu2024book,ramsaysilv2005,ramsayhook2009}:
\begin{equation}
W_i(s) = \beta_0(s) + \sum_{q=1}^Q X_{iq}\beta_q(s) + \sum_{l=1}^{L_{true}} \xi_{il} \phi_l(s) + \epsilon_i(s)\;,
\label{eq:DGM}
\end{equation}
where $\beta_q(s)$ are the fixed-effect functions of interest, $\phi_l(s)$ are the true eigenfunctions, and $L_{true}$ is the number of true eigenfunctions. The random effect scores $\xi_{il} \sim N(0, \lambda_l)$ and the error term $\epsilon_{i}(s) \sim N(0, \sigma_\epsilon^2)$ are assumed to be mutually independent, while $\epsilon_{i}(s)$ are independent across $i$ and $s$. Under this data generating mechanism, the null hypothesis of no association between covariate $X_{iq}$ and the functional response is:
\begin{equation}
    \mathbf{H}_{0,q}^{\rm TRUE}: \beta_q(s) = 0 \quad \text{for all } s \in S\;.
    \label{eq:true_null_hyp}
\end{equation}
The null hypothesis $\mathbf{H}_{0,q}^{\rm TRUE}$ is indexed by $q$ to indicate that it corresponds to the fixed effect $X_{iq}$ and superscripted by ``TRUE" to indicate that the hypothesis is specified under the true data generating mechanism. While here we use FoSR notation, the model is only ``functional" when smoothness of $\beta_q(s)$ is assumed over $s$. The same exact model has been used in many applications, especially in genomics and imaging, with or without explicit acknowledgment of the smooth properties of the various model components. Using RPCS method means that one first fits the misspecified model:
\begin{equation}
W_i(s) = \beta_0^M(s) + \sum_{l=1}^{L_{miss}} \xi^M_{il} \phi^M_l(s) + \epsilon^M_i(s)\;,
\label{eq:misspecified}
\end{equation}
and then regresses $\xi^M_{il}$ on $X_{iq}$ using the standard linear models $\xi^M_{il}=b_{0l} + \sum\limits_{q=1}^{Q}b_{ql}X_{iq}+e_{il}$, for $l=1,\ldots,L_{miss}$ where $e_{il}\sim N(0,\sigma^2_q)$ are mutually independent over $i$ (study participants) and $l$ (eigenvectors).

The relationship between these two frameworks is established in the following theorem:
\begin{theorem}
\label{thm:thm1}
    Suppose that $W_i(s)$ is generated by the true data-generating mechanism in model~\eqref{eq:DGM}. 
    Consider the regression
    $
    \xi^M_{il} = b_{0l} + \sum_{q=1}^Q b_{ql} X_{iq} + e_{il}$, $l = 1, \cdots, L_{miss}$, where the scores $\xi^M_{il}$ are estimated from the misspecified model~\eqref{eq:misspecified}. The null hypothesis for testing for no association with $X_{iq}$ is
    $$
    \mathbf{H}_{0,q}^M: b_{ql} = 0, \quad l = 1, \ldots, L_{miss}\;,
    $$
    where $b_{ql} = \int_S \beta_q(s) \phi_l^M(s) \, ds$, the inner product between $\beta_q(s)$ and $\phi_l^M(s)$.
\end{theorem}

We briefly illustrate the implementation of these two modeling strategies. While Equation 1 formalizes the True Model (FoSR) and Equation 3 outlines the Misspecified Model (RPCS), fitting them in practice requires fundamentally different software syntax. The following R code snippet demonstrates this contrast: the RPCS approach extracts discrete principal component scores prior to standard linear regression, whereas the FoSR approach jointly estimates the continuous functional fixed effects and subject-specific random effects within a single generalized additive model.
\lstset{
  language=R,
  basicstyle=\ttfamily\small,
  numbers=left,
  numberstyle=\tiny,
  frame=lines,
  framesep=2mm,
  breaklines=true,
  commentstyle=\color{gray}
}
\begin{lstlisting}
# ---------------------------------------------------------
# 1. RPCS Implementation (Misspecified Model)
# ---------------------------------------------------------
# Extract FPCA scores and regress each component on the covariate
fpca <- fpca.face(Y = Ymat, npc = L, argvals = seq(0, 1, length.out = K))
# (Merged with covariates into 'scores' dataframe)
# Example for a single Principal Component and Covariate:
fit_rpcs <- lm(PC1 ~ X1, data = scores) 
pval_rpcs <- summary(fit_rpcs)$coefficients[2, 4] 
# ---------------------------------------------------------
# 2. FoSR Implementation (True Model)
# ---------------------------------------------------------
# Fit a GAM using smooth terms for covariates and random effects for eigenfunctions
# s(time, by = X1) represents the functional fixed effect
# s(id, by = phi1) represents the subject-specific random effect scores
gam_formula <- 
as.formula("Y ~ s(time, by = X1, bs = 'cr', k = 30) + 
                                s(id, by = phi1, bs = 're')")
gam_fit <- bam(gam_formula, method = "fREML", data = df_func, discrete = TRUE)
\end{lstlisting}

\subsection{The short story of the art of losing power}\label{subsec:power_loss}
We now investigate the practical implications of the results in Theorem~\ref{thm:thm1}, which provide insights into when and why power is lost in RPCS.
 
First, note that the null hypothesis for RPCS is formulated in terms of $b_{ql}= \int \beta_q(s) \phi_l^M(s) \, ds$ rather than the true functional effects  $\beta_q(\cdot)$. This implies that if the true effect, $\beta_q(s)$, is orthogonal to the misspecified eigenfunctions, $b_{ql} = \int\beta_q(s) \phi_l^M(s) \,ds = 0$. In this case, any test for the null hypothesis $b_{ql} = 0$ (association between the scores on the $l$-th principal component, $\phi_l^M(s)$, and the $q$-th covariate, $X_q$) will have no power, irrespective of the size or shape of $\beta_q(\cdot)$. If $\beta_q(\cdot)$ is perfectly correlated with an eigenfunction $\phi_{l^*}^M(\cdot)$, then $b_{ql^*}= \int \beta_q(s) \phi_{l^*}^M(s) \, ds=\{\int \beta^2_q(s)\, ds\}^{1/2}$ and $b_{ql}= \int \beta_q(s) \phi_{l}^M(s) \, ds = 0$ for $l \neq l^*$. In this case, testing the null hypothesis $\beta_q(s) = 0$ for every $s$ in the correctly specified model~\eqref{eq:DGM} is equivalent to testing the hypothesis $b_{ql^*} = 0$ in the misspecified model~\ref{eq:misspecified} for $l = l^*$, but has no power for $l \neq l^*$. In general, $|b_{ql}| = |\int \beta_q(s) \phi^M_l(s) ds| \leq \{\int \beta^2_q(s)ds\}^{1/2} [\int\{\phi^M_l(s)\}^2 ds]^{1/2} = \{\int \beta^2_q(s)ds\}^{1/2}$, which is the $L^2$ norm of $\beta_q(\cdot)$. Therefore, for each $l$, 
the power of any test is essentially controlled by the $L^2$ correlation between the true fixed effects parameter $\beta_q(s)$ and the eigenfunctions $\phi^M_l(s)$. RPCS replaces testing for $\beta_q(\cdot) = 0$ with $L_{miss}$ tests, each with a smaller signal than the one in the data. Replacing testing for the true effect  using FoSR with testing for many hypotheses with smaller signals can potentially lead to substantial power losses. Another problem could be that the tests in the misspecified model need to be controlled for multiplicity to maintain the Family Wise Error Rate (FWER) error rate. Correctly accounting for multiplicity would further reduce the power of RPCS.
 
Second, $b_{ql}$ is difficult to interpret in the original data space, especially when the number of covariates, $q$, and PCs ($L_{miss}$) increases. Indeed, $b_{ql}$ is the projection of the true effect, $\beta_{q}(\cdot)$, on the $l$th eigenvector $\phi^M_l(\cdot)$. To make things concrete, consider the case when $L_{miss} = 3$ and one tests for $b_{ql} = 0$ and the first and third hypothesis are rejected, but the second is not (at some predetermined $\alpha$ level). In practice, choosing between an estimator that includes only significant components, $\widehat{\beta}_q(s)=\widehat{b}_{q1}\phi_1^M(s)+\widehat{b}_{q3}\phi_3^M(s)$, or one that includes all components, $\widehat{\beta}_q(s)=\widehat{b}_{q1}\phi_1^M(s)+\widehat{b}_{q2}\phi_2^M(s)+\widehat{b}_{q3}\phi_3^M(s)$, leads to fundamentally different functional shapes and interpretations. 

Third, size and power of the tests for $b_{ql} = 0$ are difficult to control, as the number of hypotheses, $Q \times L_{miss}$, grows rapidly, and the correlation structure among the test statistics is highly complex. We are not aware of any applications that explicitly acknowledge and accounts for these correlations. The problem is amplified when covariates are selected or when models are used to adjust progressively for covariates. This oversight causes a significant risk in data analysis, as it may lead to false negatives or biased estimators of true biological effects.

We now demonstrate through simulations and real-data applications that these theoretical insights correspond exactly to the practice of data analysis. In Section \ref{sec:simulations}, we investigate the power and size for testing fixed effects under various synthetic data generating mechanisms scenarios. In Section \ref{sec:real_data}, we show the real-world consequences of this power loss using minute-level physical activity profiles from the 2011–2014 NHANES, across different age groups. 

\section{Simulations}\label{sec:simulations}
We conducted an extensive simulation study by generating functional data based on a pre-specified Function on Scalar Regression (FoSR) model (Equation $\eqref{eq:DGM}$). A total of $1,000$ simulations were conducted for each scenario to obtain the size and power of the tests. Two methods were compared: (1) FoSR using explicit modeling of the dependence between the fixed effects and the principal components (the True Model, TM); and (2) the regression of principal component scores (RPCS) on covariates (the Misspecified Model, MM). The specific null hypotheses were defined as: $\mathbf{H}_{0,q}^{\rm TRUE}: \beta_q(s) = 0$ for all $s$ (no association between the $q$-th scalar predictor and the functional response) for the TM test, and $\mathbf{H}_{0,q}^M: b_{ql} = 0$ for all $l = 1, \ldots, L_{miss}$ (no association between the $q$-th scalar predictor and any of the $L_{miss}$ estimated FPCA scores) for the MM test. 

\subsection{Scenario 1: One Fixed Effect and One Principal Component}\label{subsec:scenario1}

The first scenario, labeled (TM1), considers the simplest data generating mechanism
$$\text{(TM1)}\ \ W_i(s) = X_{i1}\beta_{1,d,w}(s) + \xi_{i1}\phi_1(s) + \varepsilon_{i}(s)\;,$$
Here, the scalar predictors $X_{i1}$ are independent realizations of ${\rm Uniform}(0, 1)$ random variable. The functional random effect is $\xi_{i1}\phi_1(s)$, where $\phi_1(s) = \sqrt{2}\sin(2\pi s)$ and the scores $\xi_{i1}$ are independent $N(0, \lambda_1)$ random variables with $\lambda_1 = 0.5$. The functional coefficient $\beta_{1,d,w}(s) = d \{w \times 1 + (1-w) \times \sqrt{2}\sin(2\pi s)\}$ is indexed by the parameters $d \in [0,1]$ and $w \in [0,1]$. The parameter $d$ controls the effect size, where $d=0$ corresponds to $\beta_{1,d,w}(s)=0$, or zero effect. The parameter $w$ controls the correlation between $\beta_{1,d,w}(s)$ and the principal component $\phi_1(s)$, as $\int_0^1 \beta_{1,d,w}(s) \phi_1(s)ds = (1 - w)d$ and thus ${\rm Cor}\{\beta_{1,d,w}(s),\phi_1(s)\}=(1-w)/\sqrt{w^2+(1-w)^2}$. When $w = 0$, the fixed effect $\beta_{1,d,w}(s)$ is perfectly collinear with $\phi_1(s)$ and when $w=1$, it is orthogonal. By varying $w$ between $0$ and $1$ (while holding $d$ fixed), the correlation between $\beta_{1,d,w}(s)$ and $\phi_1(s)$ varies smoothly from $1$ to $0$. 

The null hypothesis for the true model is $\mathbf{H}_0^{\rm TRUE}: \beta_1(s) = 0 \text{ for all } s \in S$. To examine the consequence of using RPCS, we consider the misspecified model (MM1):
$$\text{(MM1)}\ \ W_i(s) = \xi_{i}^M \phi_1^M(s) + \epsilon_i^M (s)\;,$$
which omits the fixed effect. The estimated scores $\xi_i^M$ are then regressed on the scalar predictor, $\xi_i^M = b_0 + b_1 X_{i1} + e_i$, and inference is conducted on $\mathbf{H}_0^{M}: b_1 = 0$. 

\begin{figure}[ht]
\centering
\begin{tabular}{|c|c|c|}
\hline
$w$ & $\beta_{1, d, w}(x)$ \includegraphics[width=5cm]{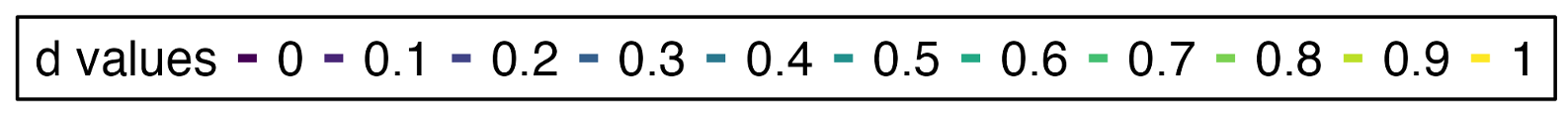} & Power of $\beta_{1, d, w}(s)$ \\
\hline
0  & \includegraphics[width=4.8cm]{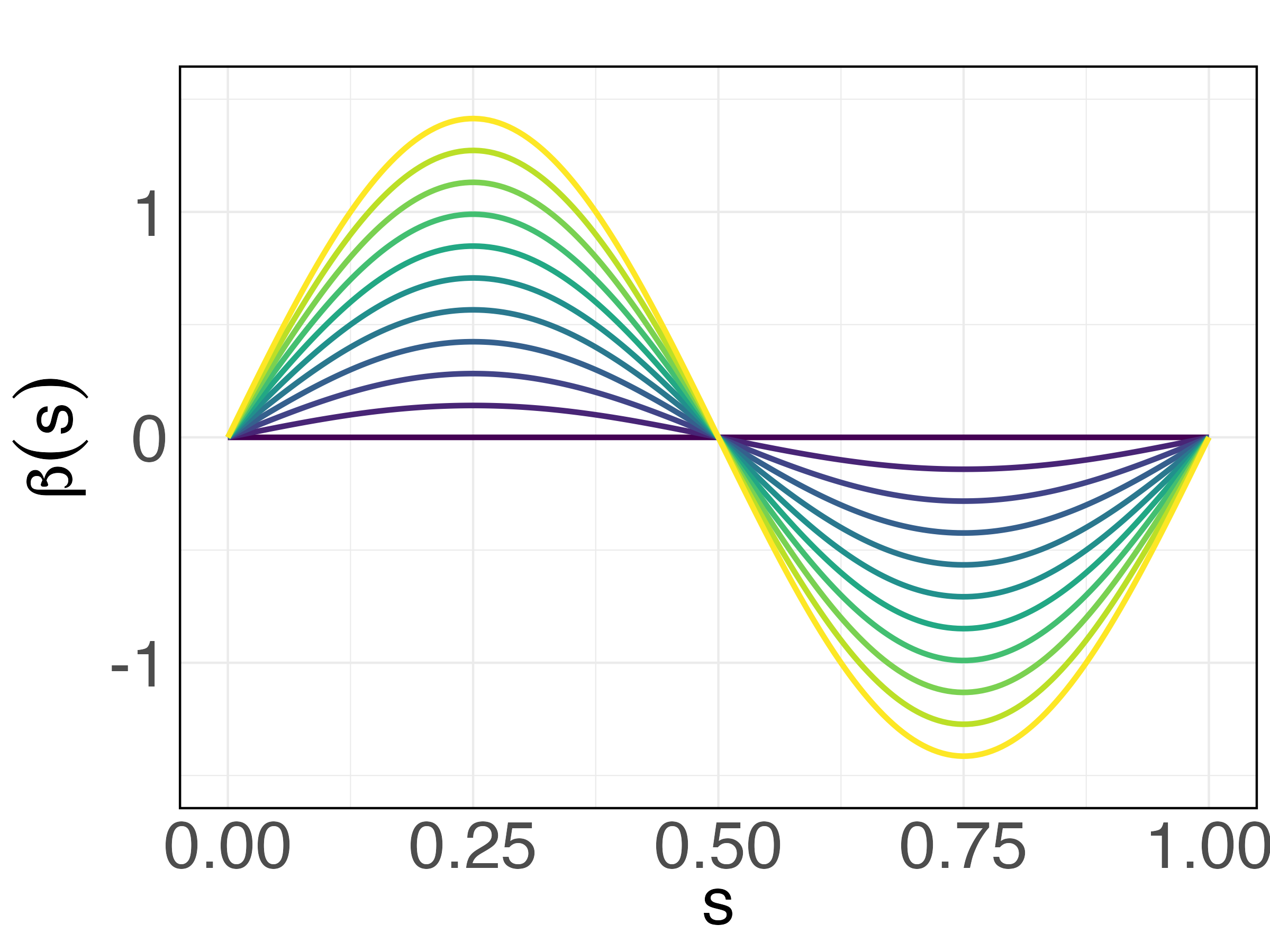} 
   & \includegraphics[width=4.8cm]{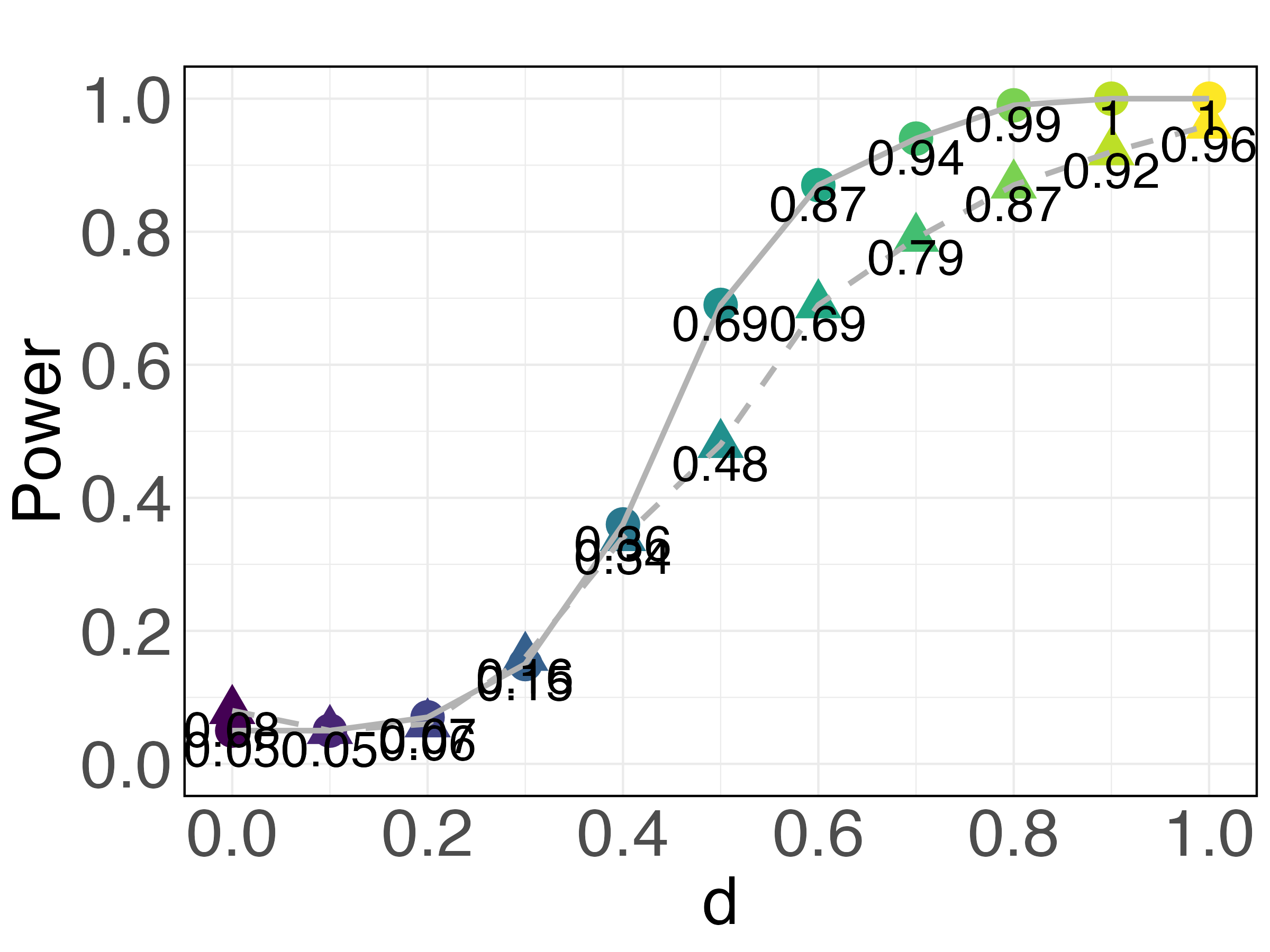} \\[2mm]

0.5 & \includegraphics[width=4.8cm]{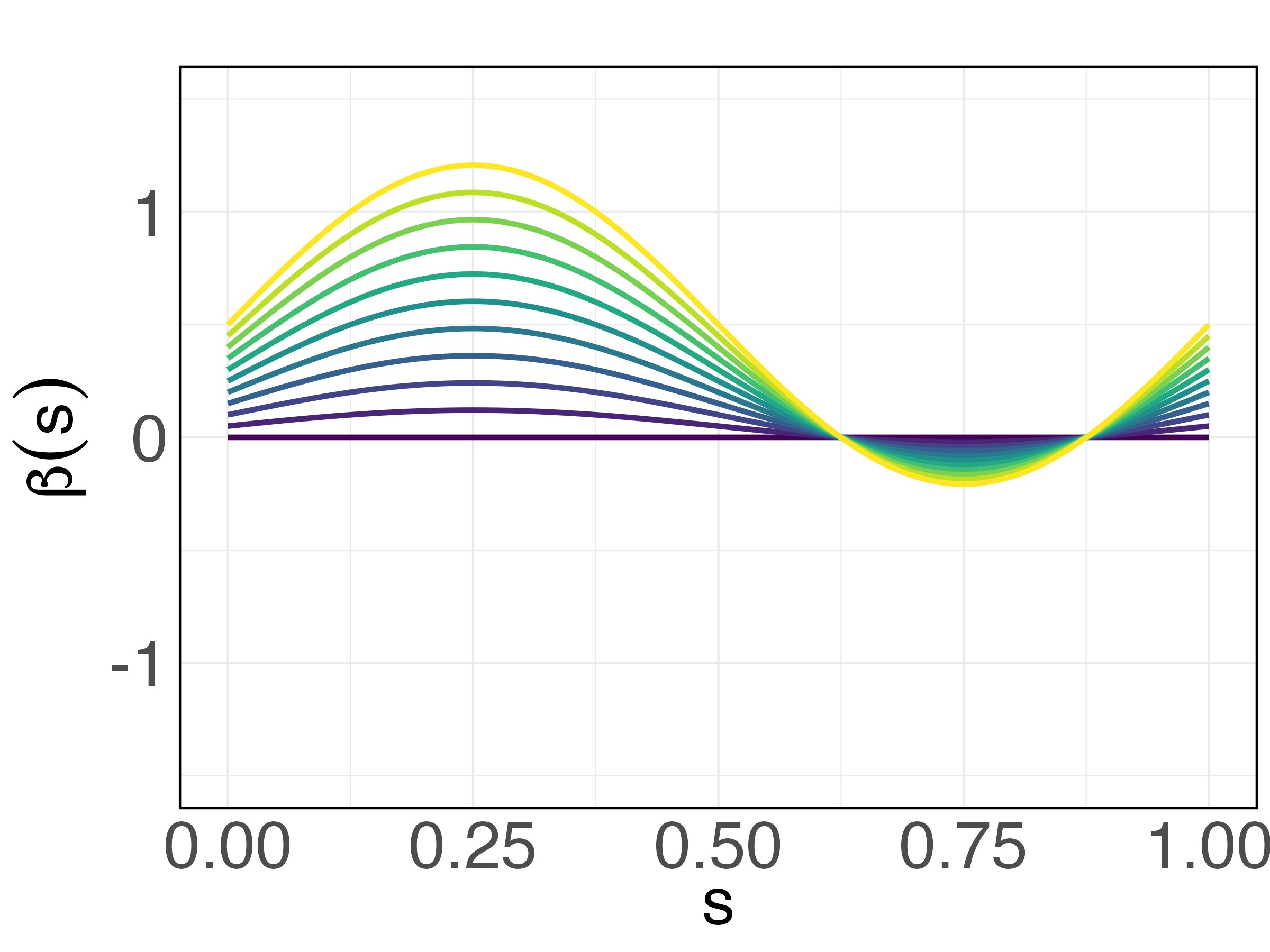} 
   & \includegraphics[width=4.8cm]{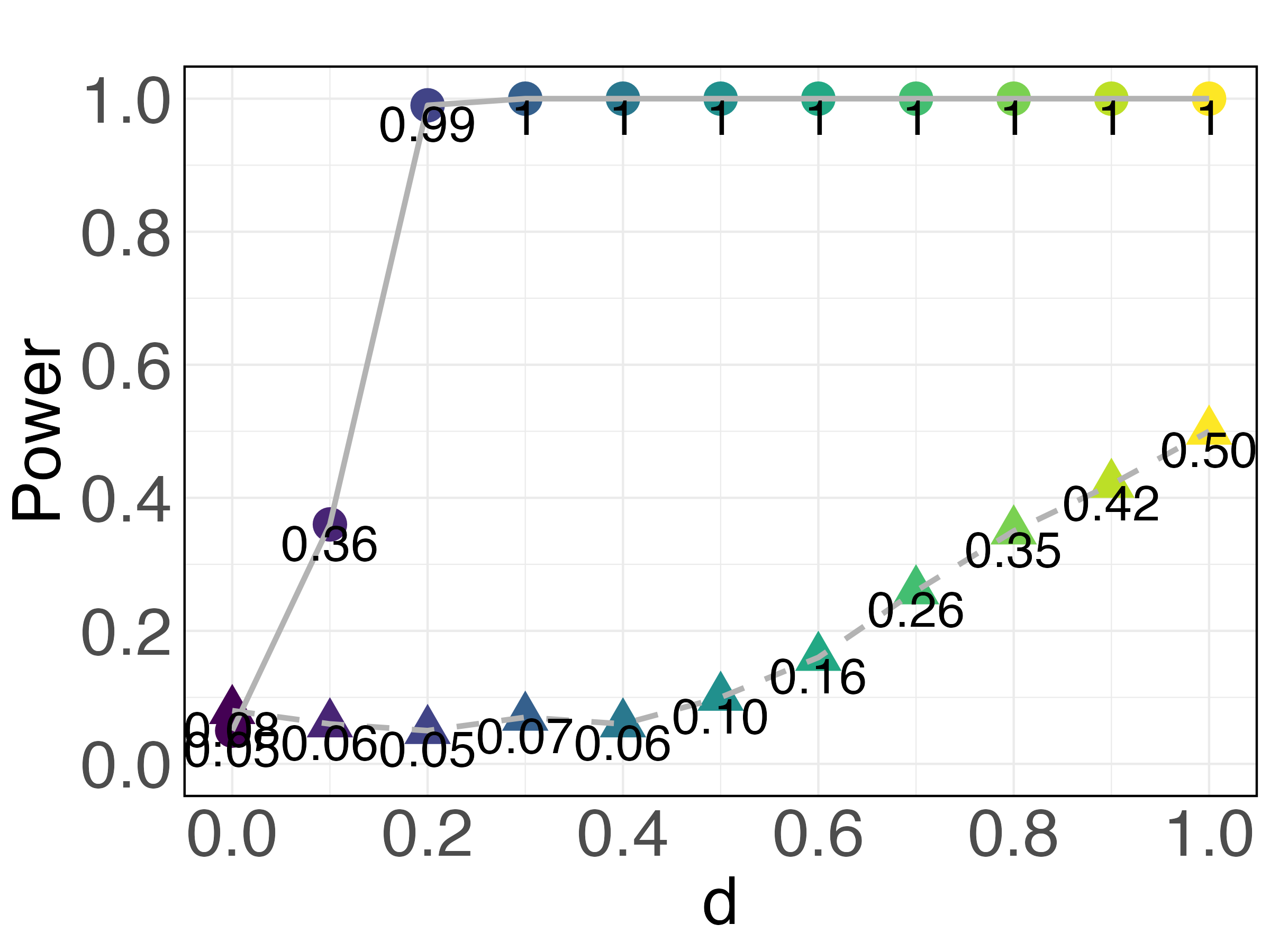}\\[2mm]

1.0 & \includegraphics[width=4.8cm]{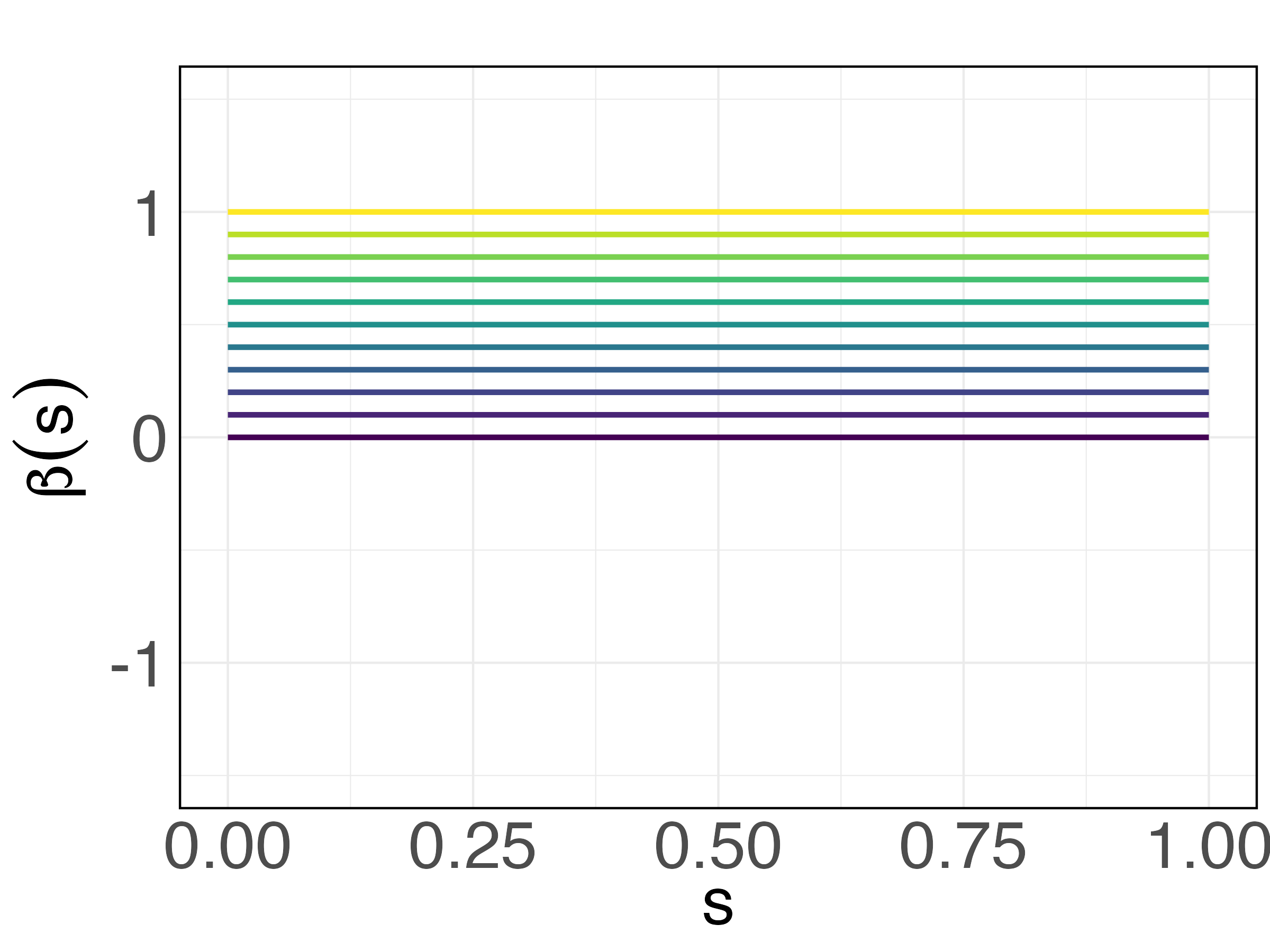} 
   & \includegraphics[width=4.8cm]{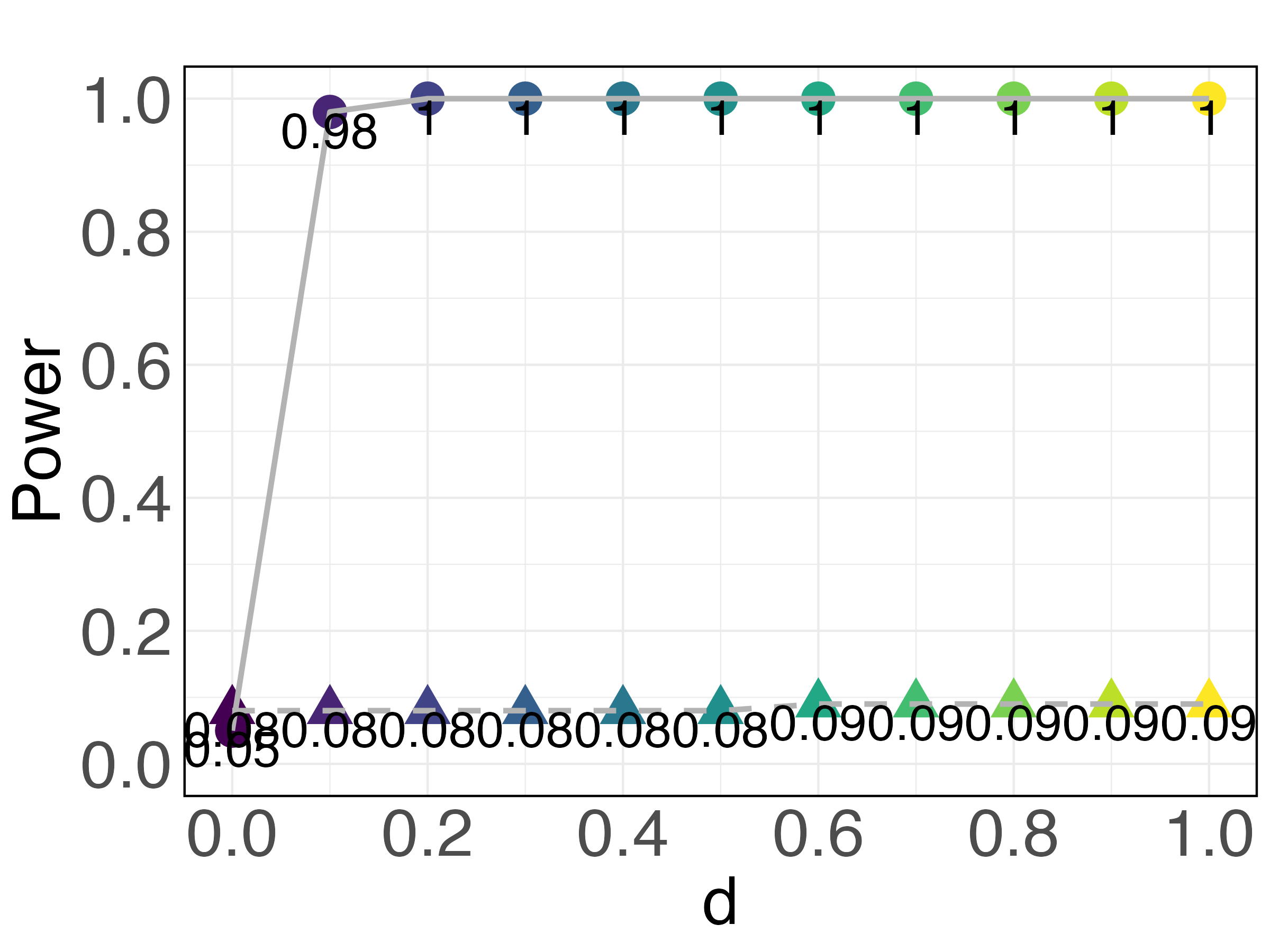}\\[2mm]
\hline
\end{tabular}
\caption{Power of FoSR (solid line in the third column) and RPCS (dashed line in the third column) for Scenario 1 for different covariance ($w$, shown by rows) and effect size ($d$, shown by different colors in the second and third column).}
\label{fig:simu_1b1p_dw}
\end{figure}

Figure \ref{fig:simu_1b1p_dw} summarizes the empirical results of Scenario 1 across different combinations of effect sizes, $d$, and correlation parameters, $w$. Each row corresponds to a specific value of $w$ listed in the first column, with lower values of $w$ (top) corresponding to lower correlation between the fixed effect, $\beta_{1,d,w}(s)$, and the principal component, $\phi_1(s)$. The second column displays the functional coefficients $\beta_{1,d,w}(s)$ for $d$ ranging from $0$ (dark purple) to $1$ (light yellow). The third column presents the empirical power for detecting if $\beta_{1,d,w}(s)$ is different from 0 for the correctly specified FoSR model (TM1, power curve shown as a solid line with round dots) and the misspecified RPCS method (MM1, power curve shown as a dashed line with triangle dots). The points on the power curve are color-coded by $d$, matching the colors used for $\beta_{1,d,w}(s)$ in the second column. The size for FoSR is close to $0.05$, while RPCS exhibits inflated size (in the $0.06$ to $0.08$ range). Despite this fact, the power of the FoSR test increases rapidly with $d$ and is always higher than the power of RPCS. These differences become more pronounced when $w$ increases (compare graphs from the first to the last row) with the RPCS power becoming very low when $w = 1$. Interestingly, the power of the FoSR approach increases with $w$ (compare the solid power line in the third column from the first to the last row), while that of the RPCS decreases. As $w$ increases the power of the RPCS quickly approaches its size (see the dashed curve in the bottom right panel with triangle dots.)  

\subsection{Scenario 2: Two Fixed Effects and Four Principal Components}\label{subsec:scenario2}
The second scenario, labeled (TM2), extends the simulation scenario in Section~\ref{subsec:scenario1} by incorporating two fixed effects and four principal components. The true data generating mechanism is:
$$
\text{(TM2) }
W_i(s) = \beta_0 + X_{i1}\beta_1(s) + X_{i2}\beta_{2,w}(s)+ \sum_{l = 1}^4 \xi_{il}\phi_l(s) + \varepsilon_{i}(s)\;,
$$ 
where $X_{i1}$ and $X_{i2}$ are mutually independent ${\rm Uniform}[0,1]$ random variables. The four principal components are defined as $\{\sqrt{2}\sin(2\pi s), \sqrt{2}\cos(2\pi s), \sqrt{2}\sin(4\pi s), \sqrt{2}\cos(4\pi s)\}$, with corresponding eigenvalues $\lambda_l = 0.5^{l-1}$ for $l = 1, \dots, 4$. 

A total of $24$ cases were considered to systematically explore the impact of various functional forms for the fixed effects $\beta_1(s)$ and $\beta_{2,w}(s)$ on the statistical power for detecting $\beta_{2,w}(s)$. Cases included: (1) $\beta_1(s)$ was either the constant function $\beta_1(s)=1$ (orthonormal to all $L=4$ PCs $\phi_l(s)$) or $\beta_1(s)=\sqrt{2}\sin(2\pi s)$ (collinear with the first PC $\phi_1(s)$); (2) the coefficient $\beta_{2,w}(s)$ is defined as a convex combination: $\beta_{2,w}(s) = w \times f_1(s) + (1 - w) \times f_2(s)$, where $f_1(s)=1$ (orthogonal to all PCs) and $f_2(s)$ is a Fourier basis function that is either among the PCs (e.g., $\sqrt{2}\sin(2\pi s)$) or not (e.g., $\sqrt{2}\sin(6\pi s)$). Just as in Section~\ref{subsec:scenario1}, as the weight parameter $w \in [0, 1]$ increase from $0$ to $1$, the function $\beta_{2,w}(s)$ smoothly transitions between the functions $f_2(s)$ (when $w = 0$) and $f_1(s)$ (when $w = 1$). We discuss results for the first $6$ cases, while the supplementary materials contain results for an additional $18$ cases. The misspecified model in this case is 
$\text{(MM) }Y_i(s) = \sum_{l = 1}^4 \xi_{il}^M\phi_l^M(s) + \epsilon_i^M (s)$.
The estimated scores $\xi_{il}^M$ are regressed on the scalar predictor, $\xi_{il}^M = b_{0l} + \sum_{q = 1}^Q b_{ql} X_{iq} + e_{il}$, and inference is conducted on $\mathbf{H}_{0,q}^{M}: b_{ql} = 0, l = 1, \cdots, L_{miss}$. As in Scenario 1, the functions $\beta_{1}(s)$ and $\beta_{2,w}(s)$ can be multiplied with a constant $d$, which controls the strength of the signal. When $d=0$ there is no effect and simulations indicate that both approaches have a size close to nominal. For presentation purposes, we only show results for $d=1$. 

\begin{figure}[ht]
\centering
\begin{tabular}{|c|c|c|c|}
\hline
\textbf{Case} & $\beta_1(x)$ & $\beta_2(x)$ \includegraphics[width=5cm]{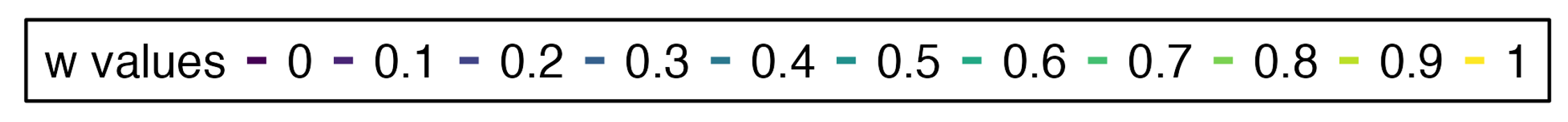} & \textbf{Power of $\beta_1(s)$, $\beta_2(s)$} \\
\hline
1 & \multirow{3}{*}{\includegraphics[width=4cm]{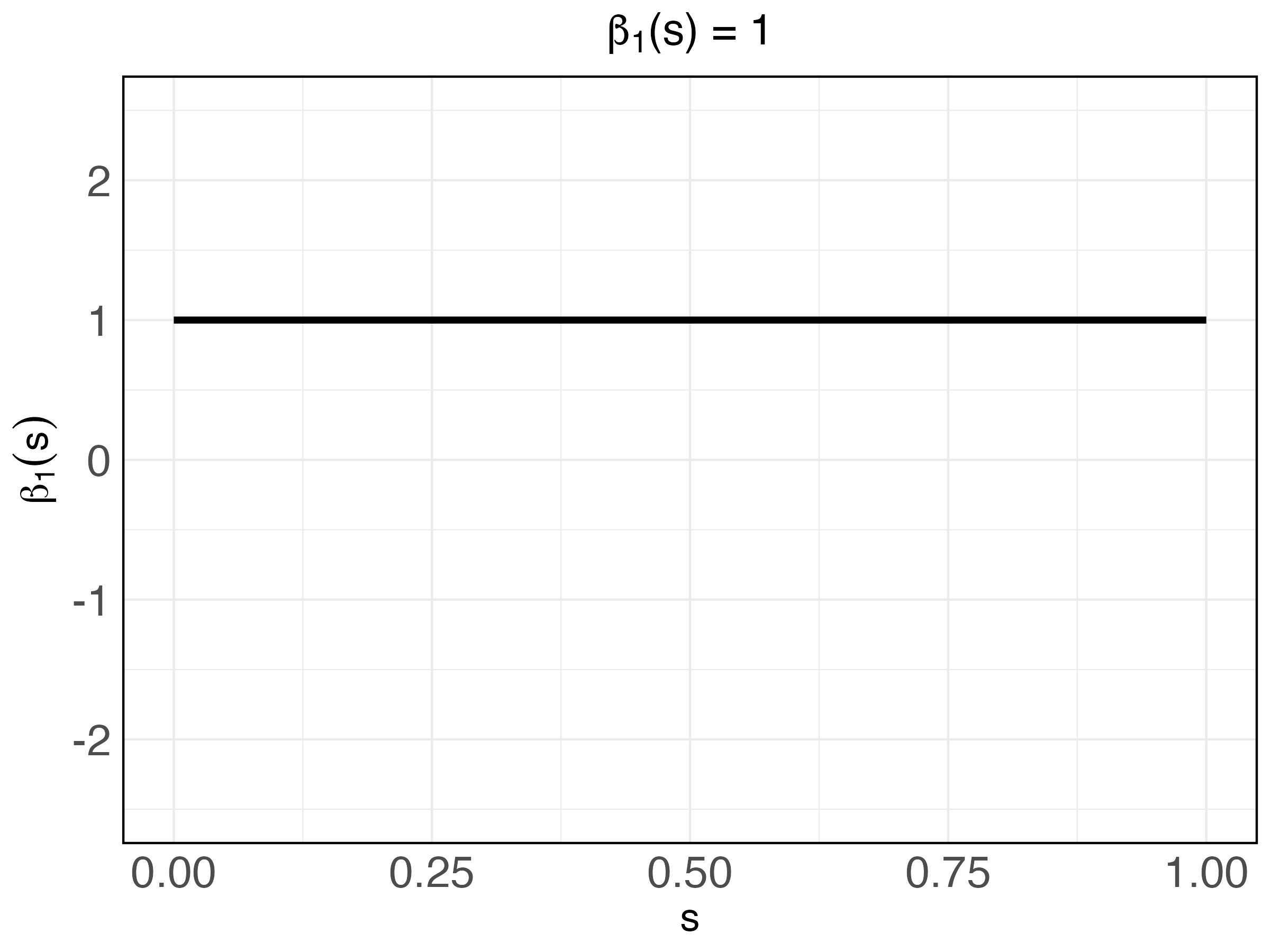}}
& \includegraphics[width=4cm]{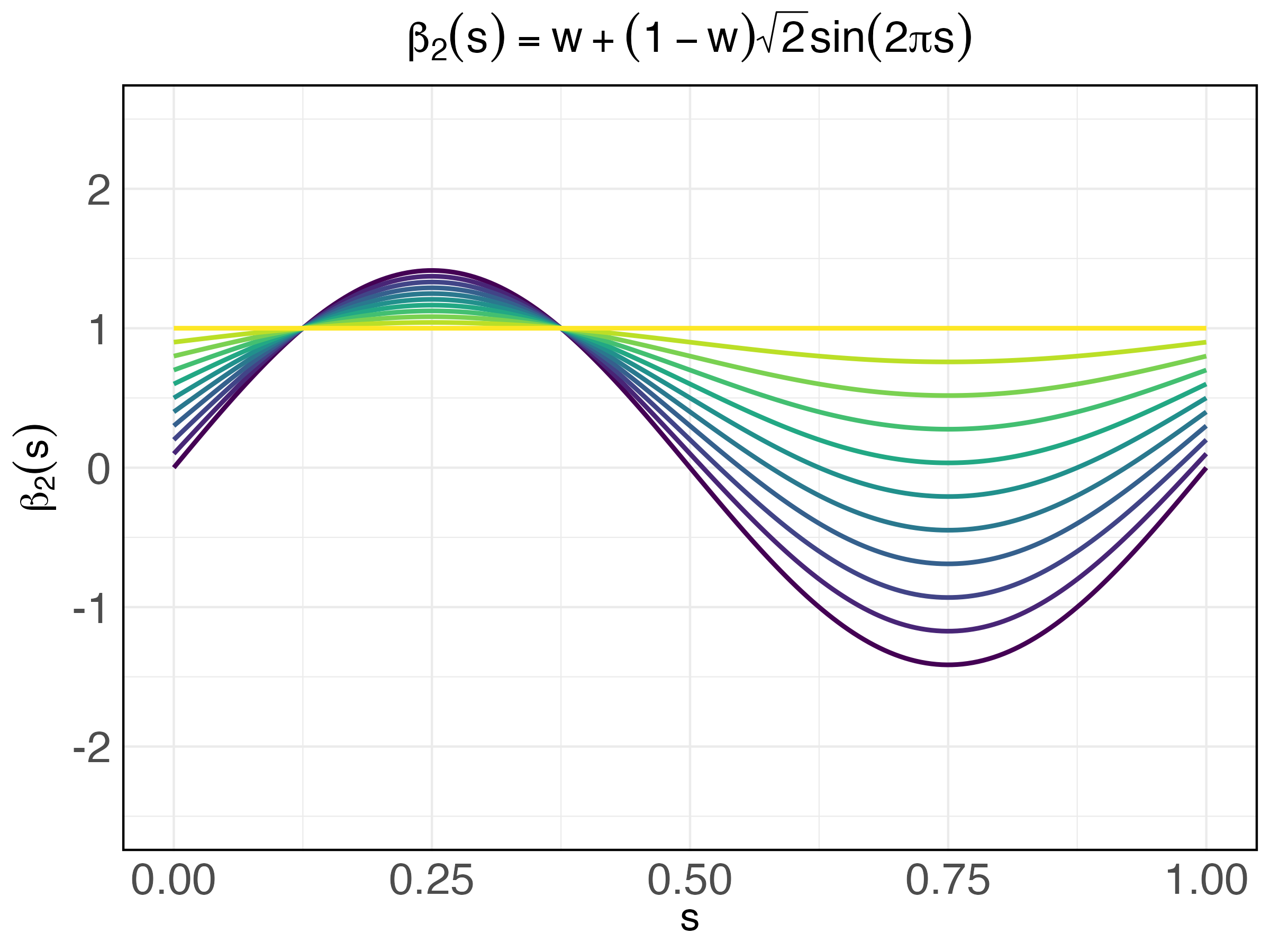}
& \includegraphics[width=4cm]{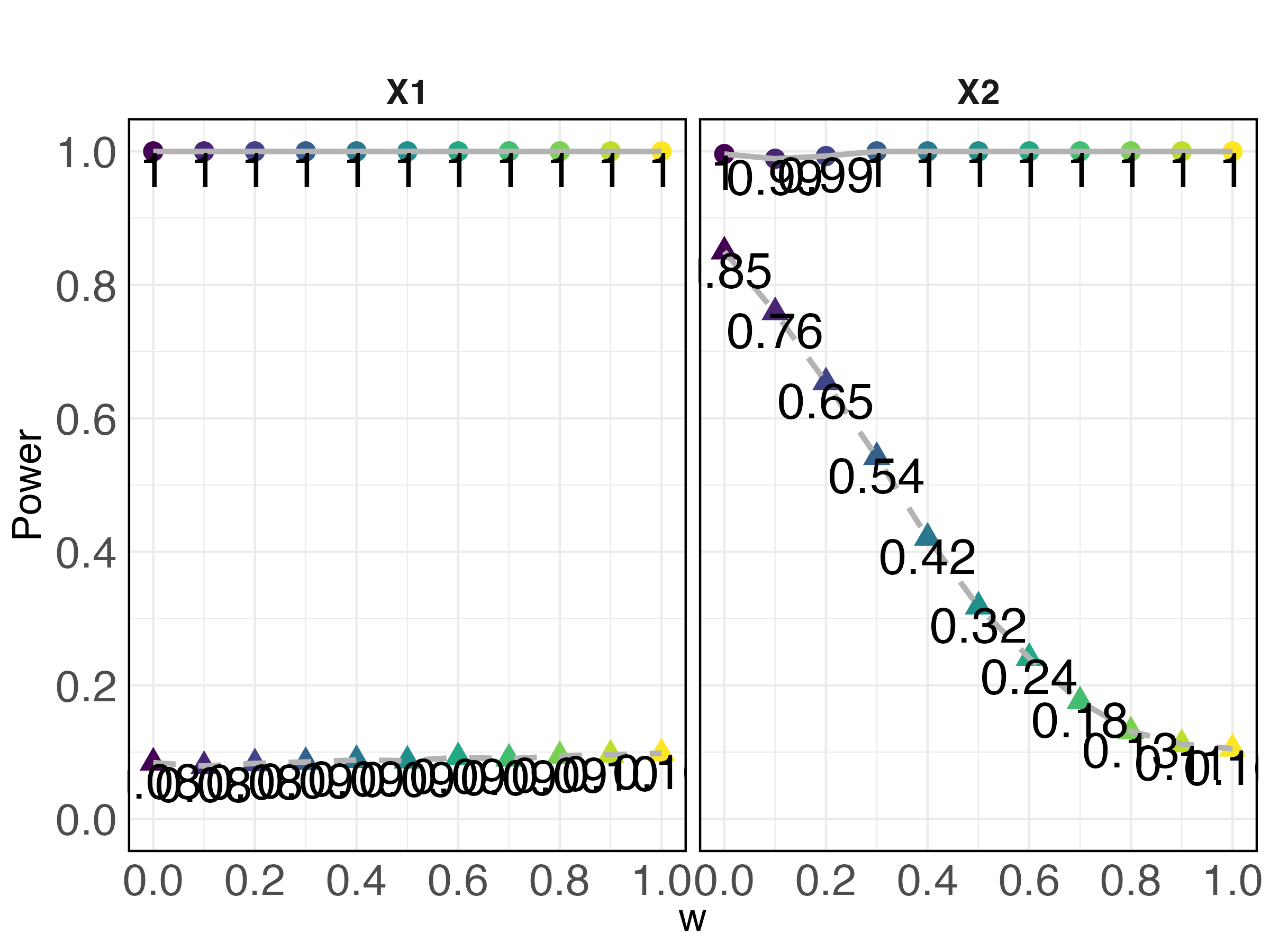} \\[2mm]
2 &
& \includegraphics[width=4cm]{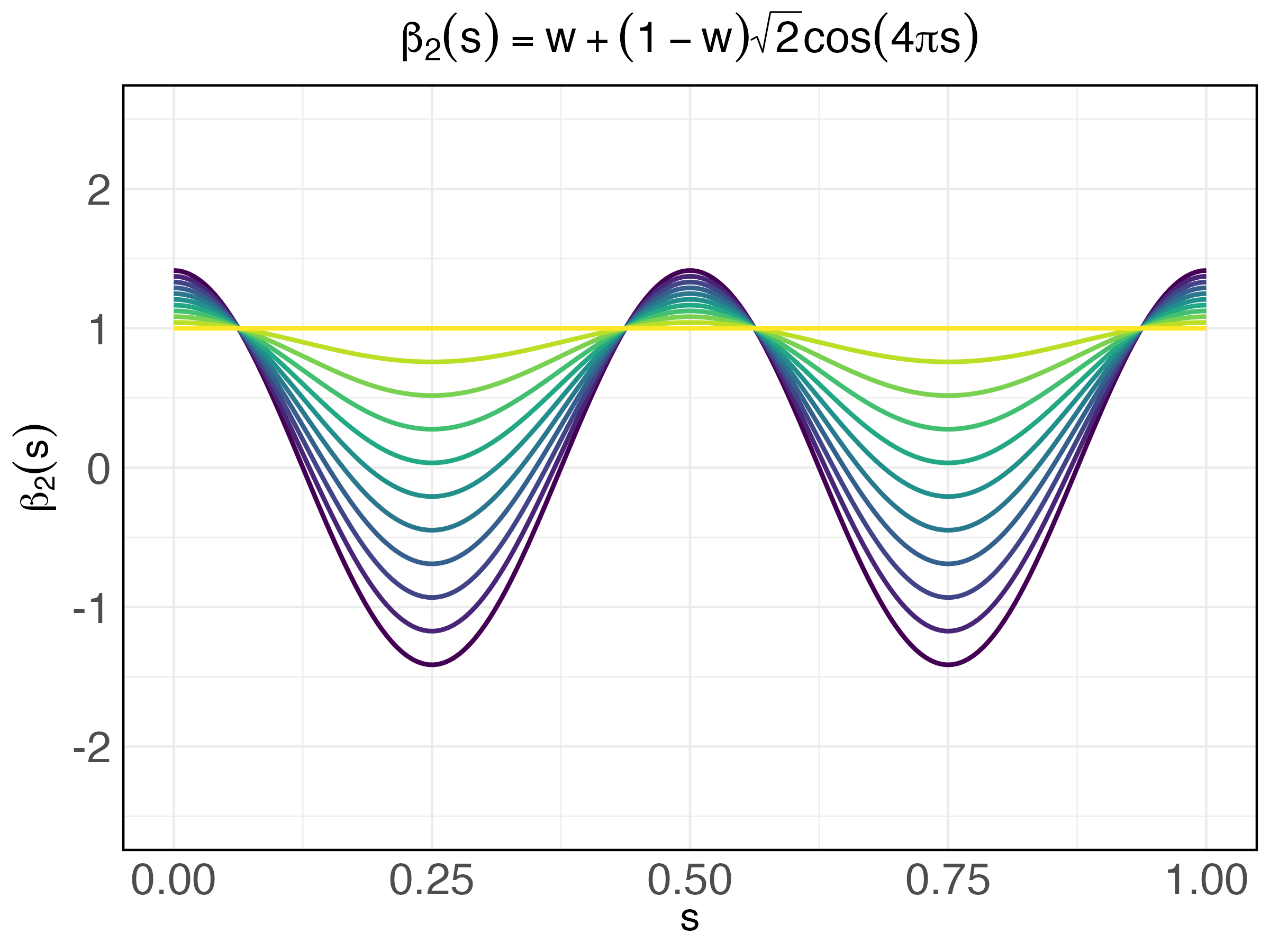}
& \includegraphics[width=4cm]{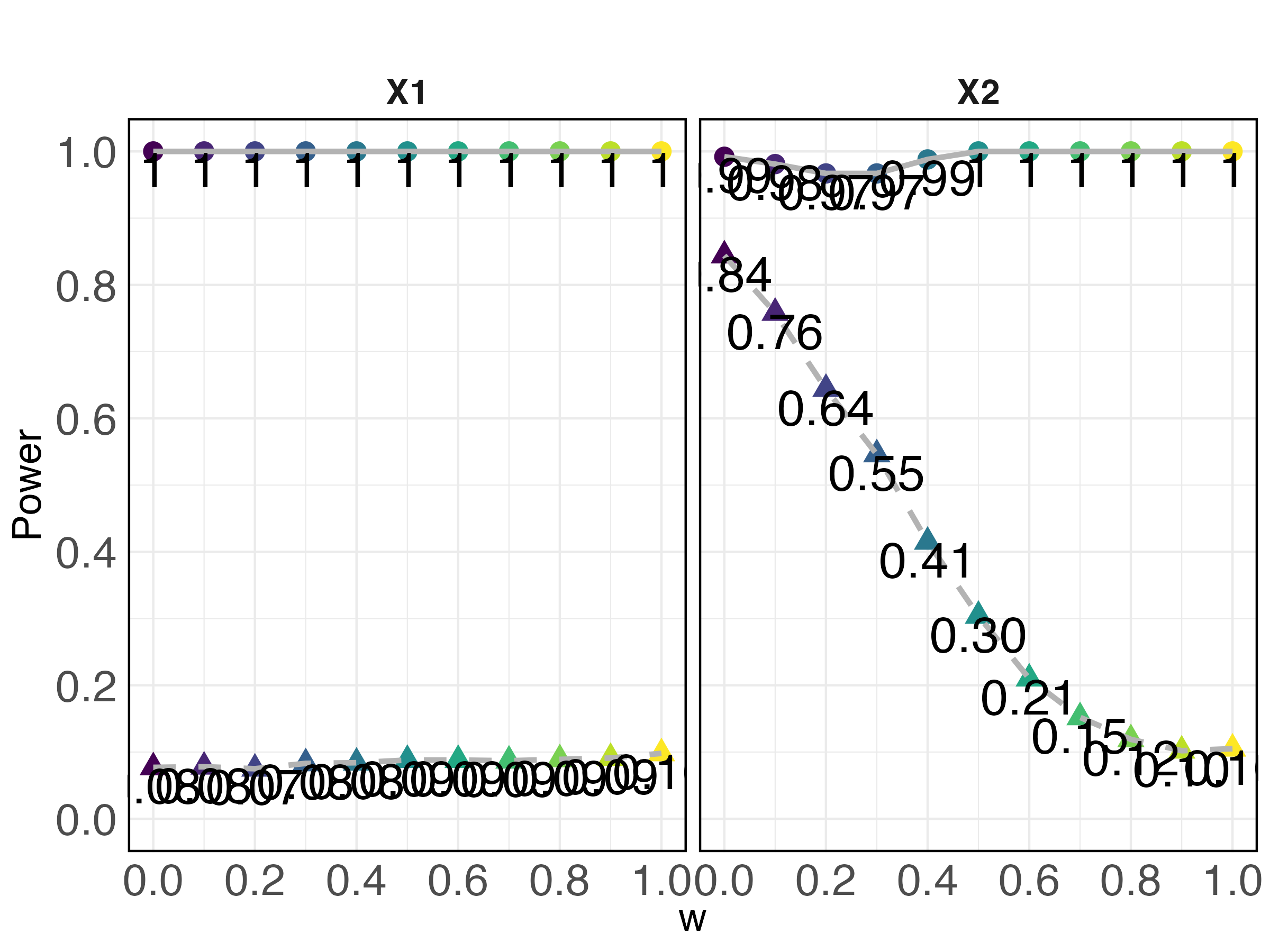} \\[2mm]
3 &
& \includegraphics[width=4cm]{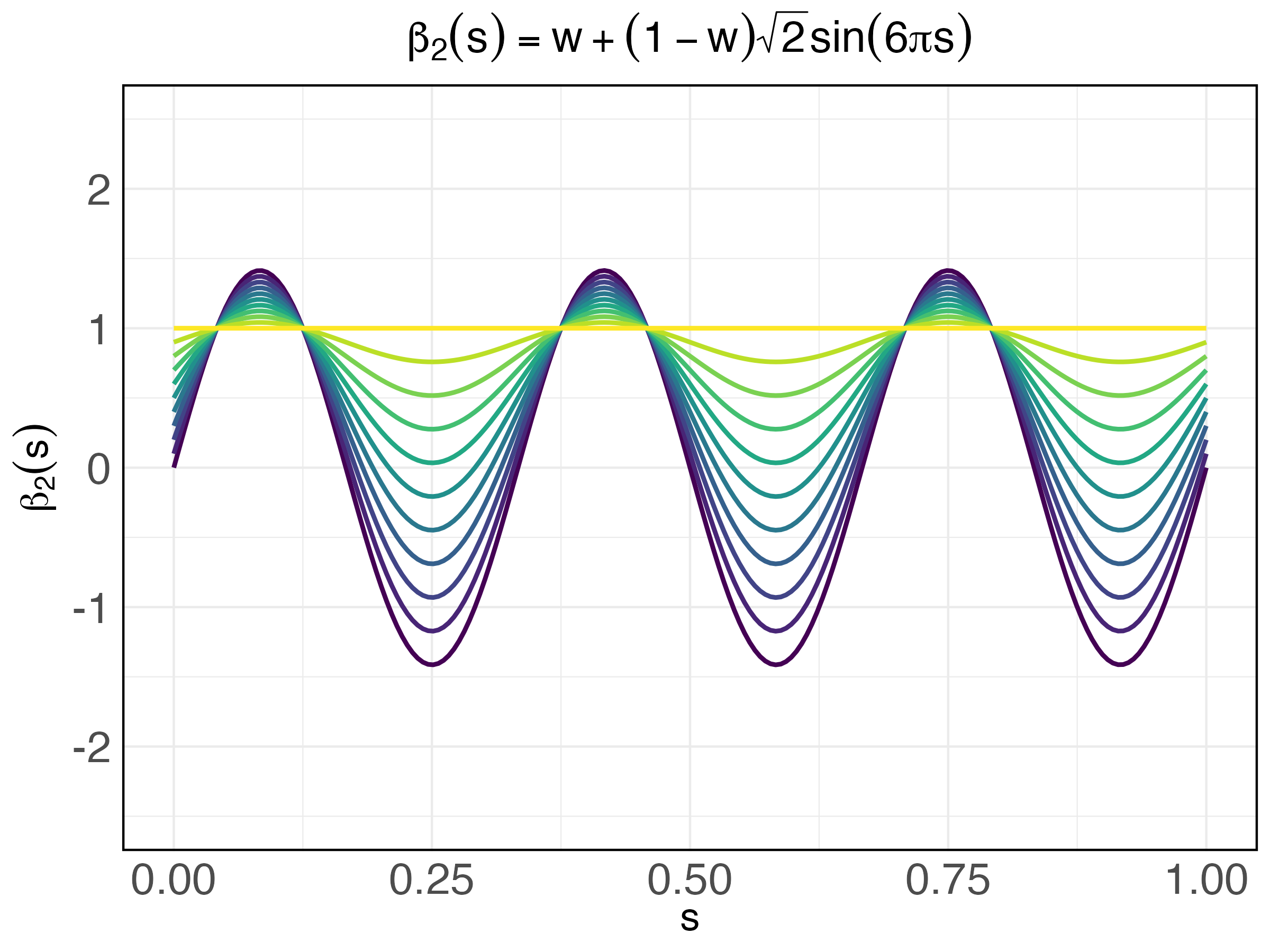}
& \includegraphics[width=4cm]{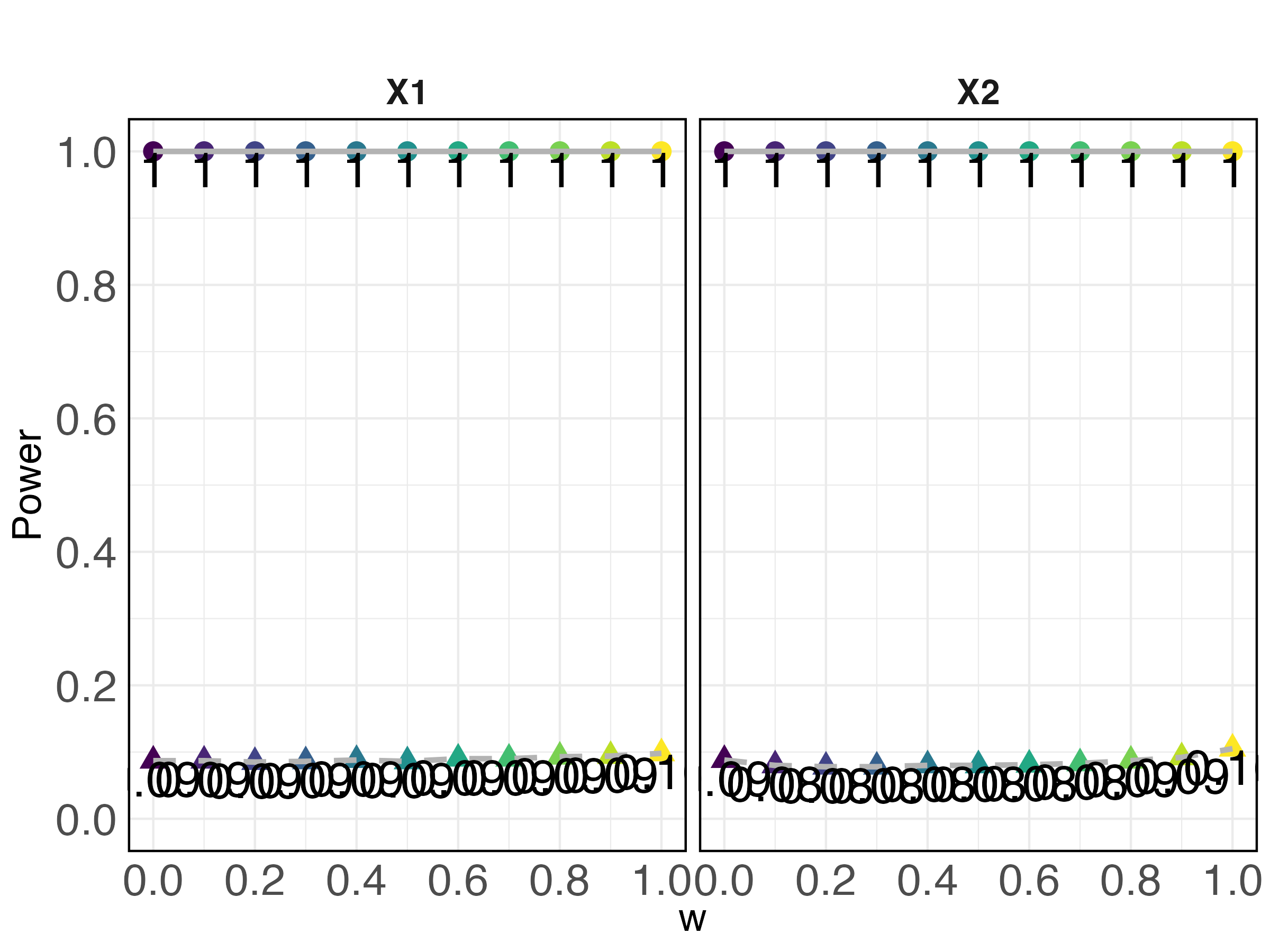} \\[2mm]
\hline
4 & \multirow{3}{*}{\includegraphics[width=4cm]{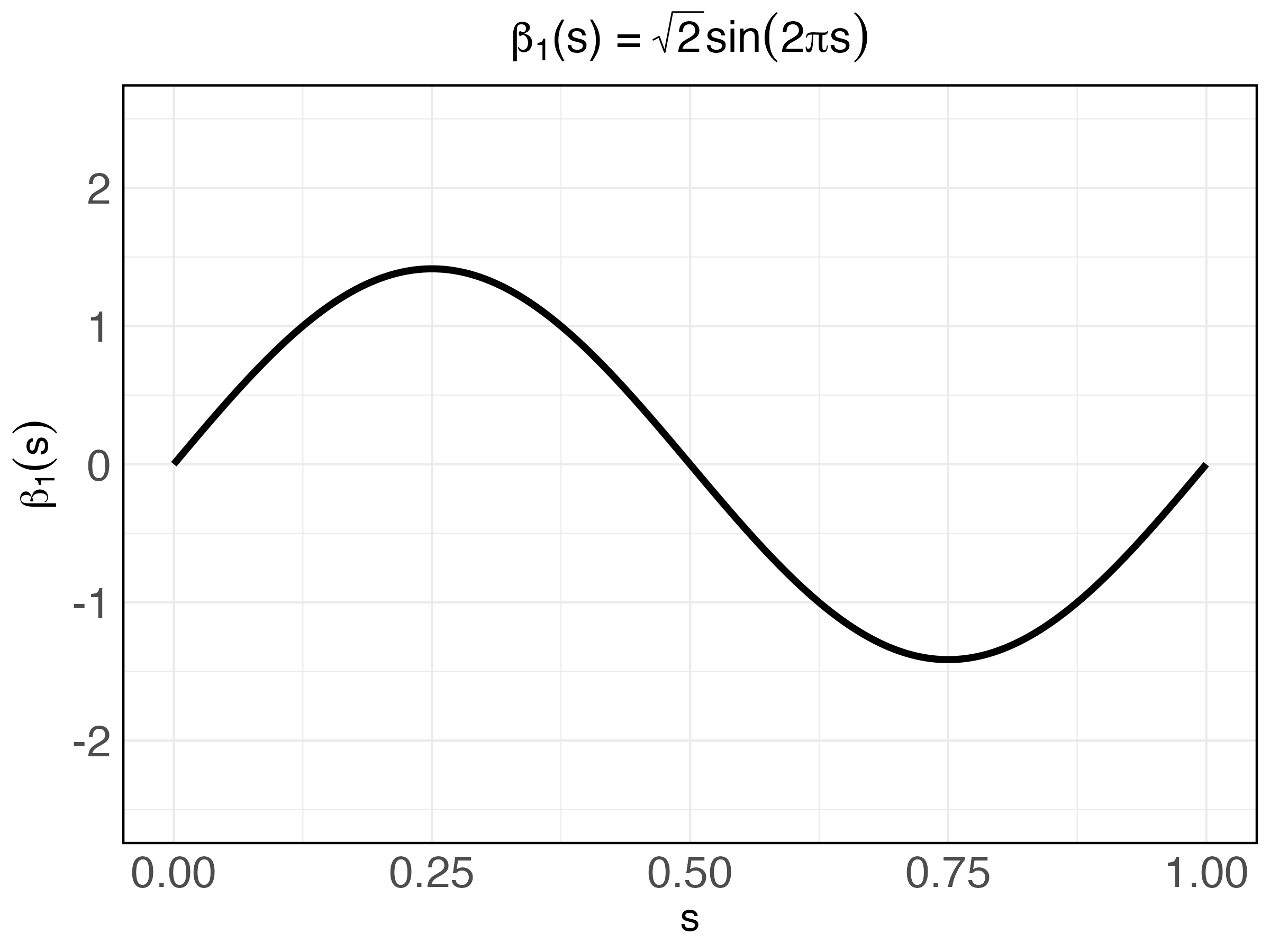}}
& \includegraphics[width=4cm]{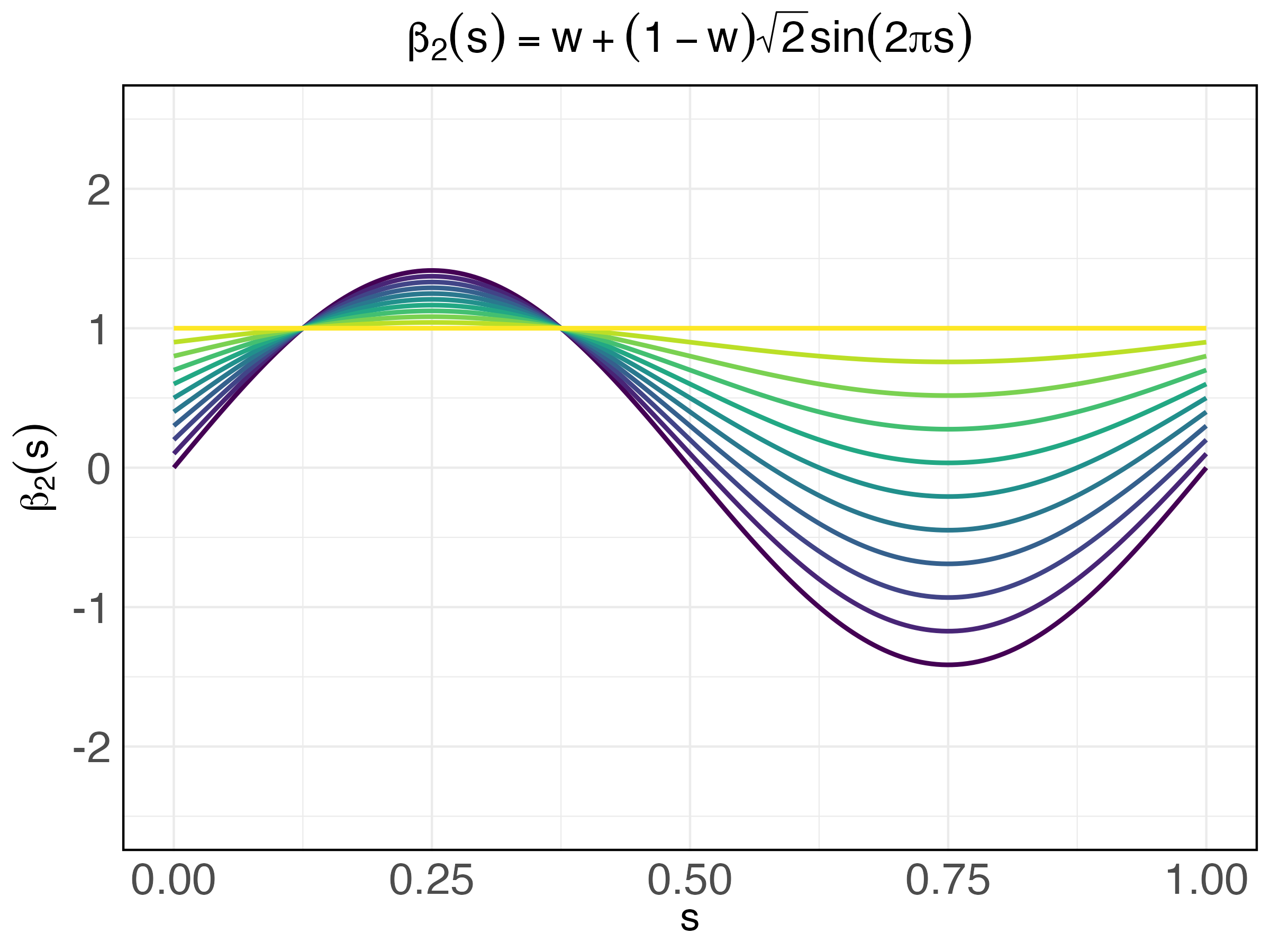}
& \includegraphics[width=4cm]{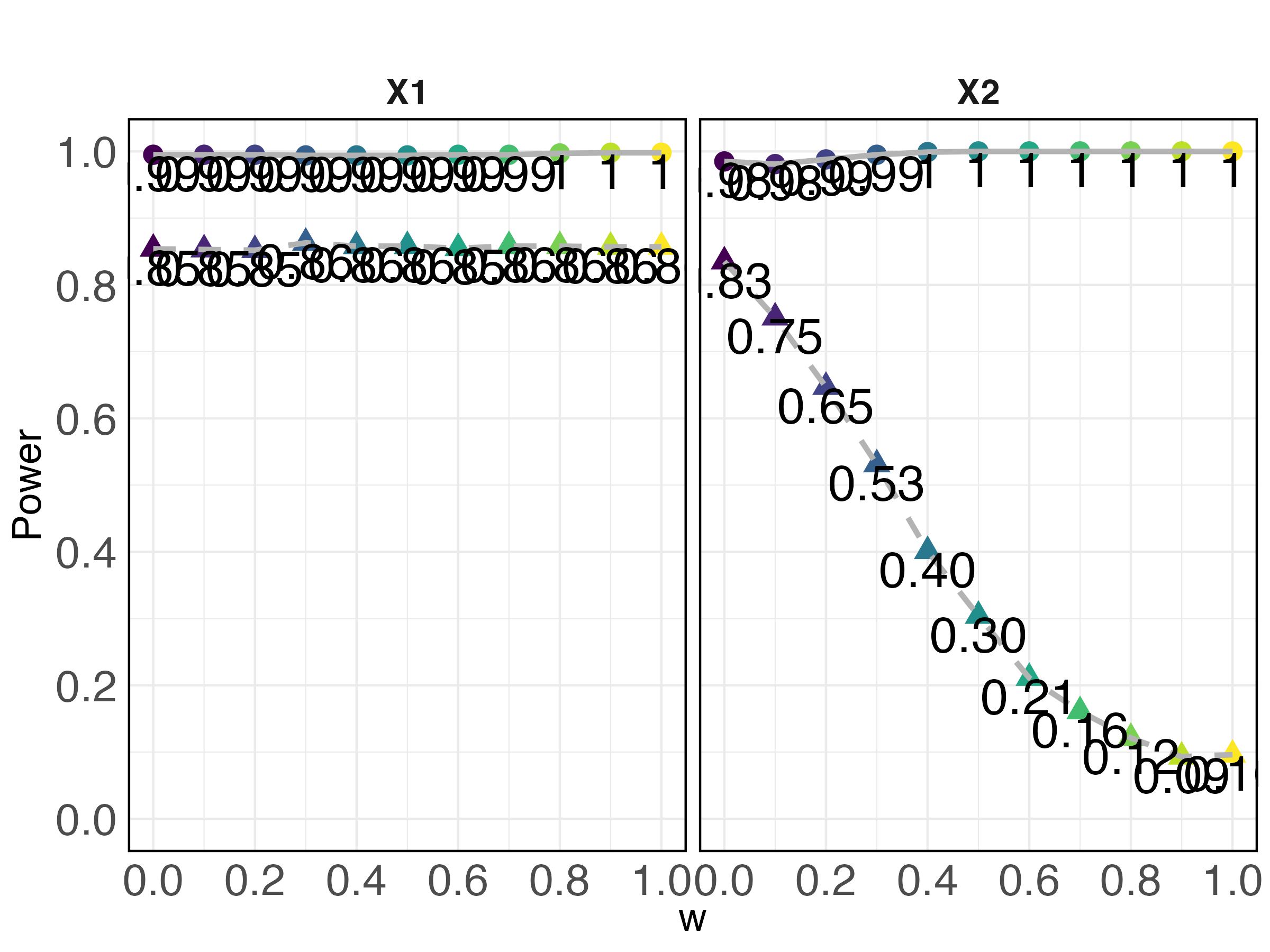} \\[2mm]
5 &
& \includegraphics[width=4cm]{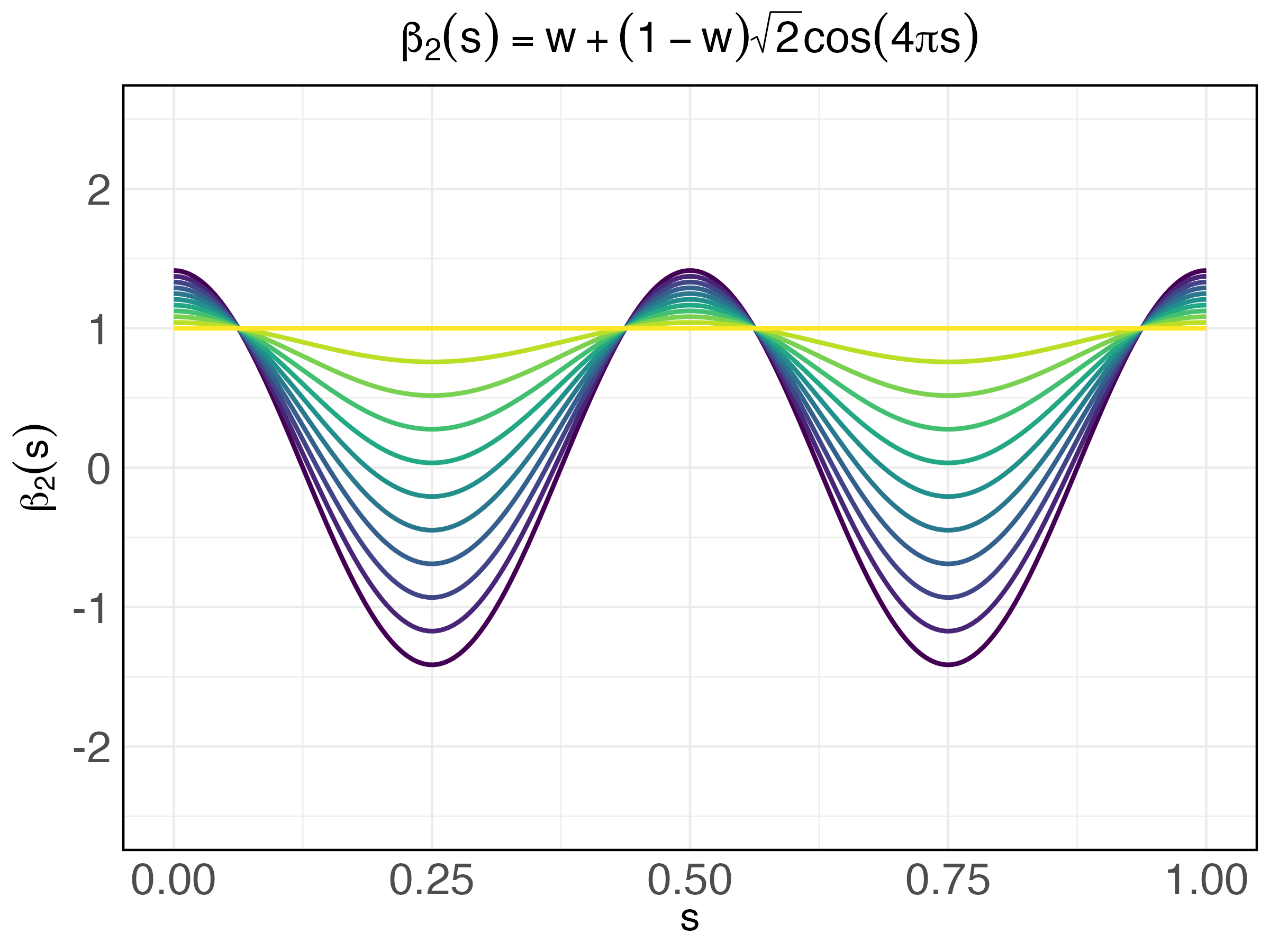}
& \includegraphics[width=4cm]{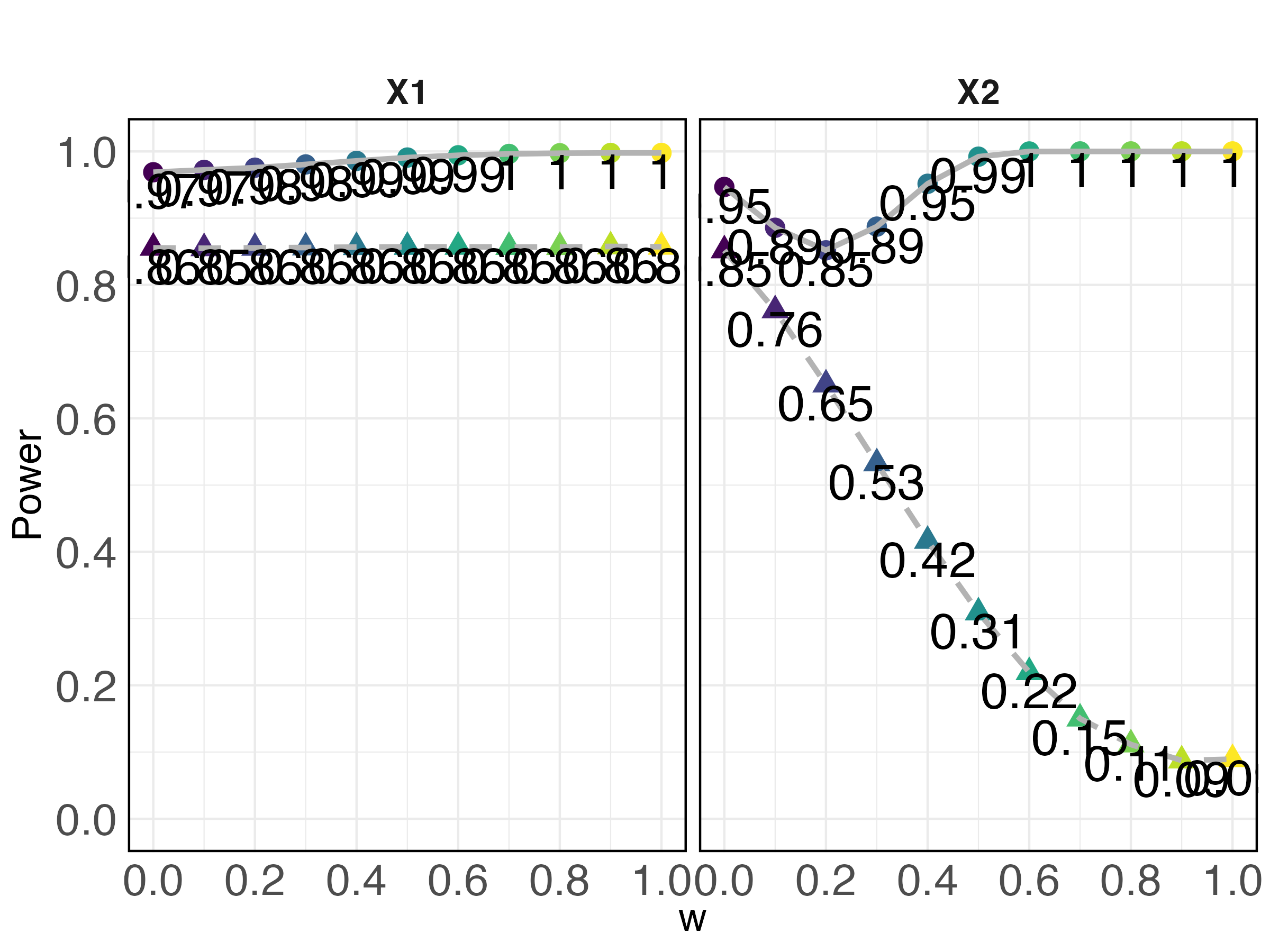} \\[2mm]
6 &
& \includegraphics[width=4cm]{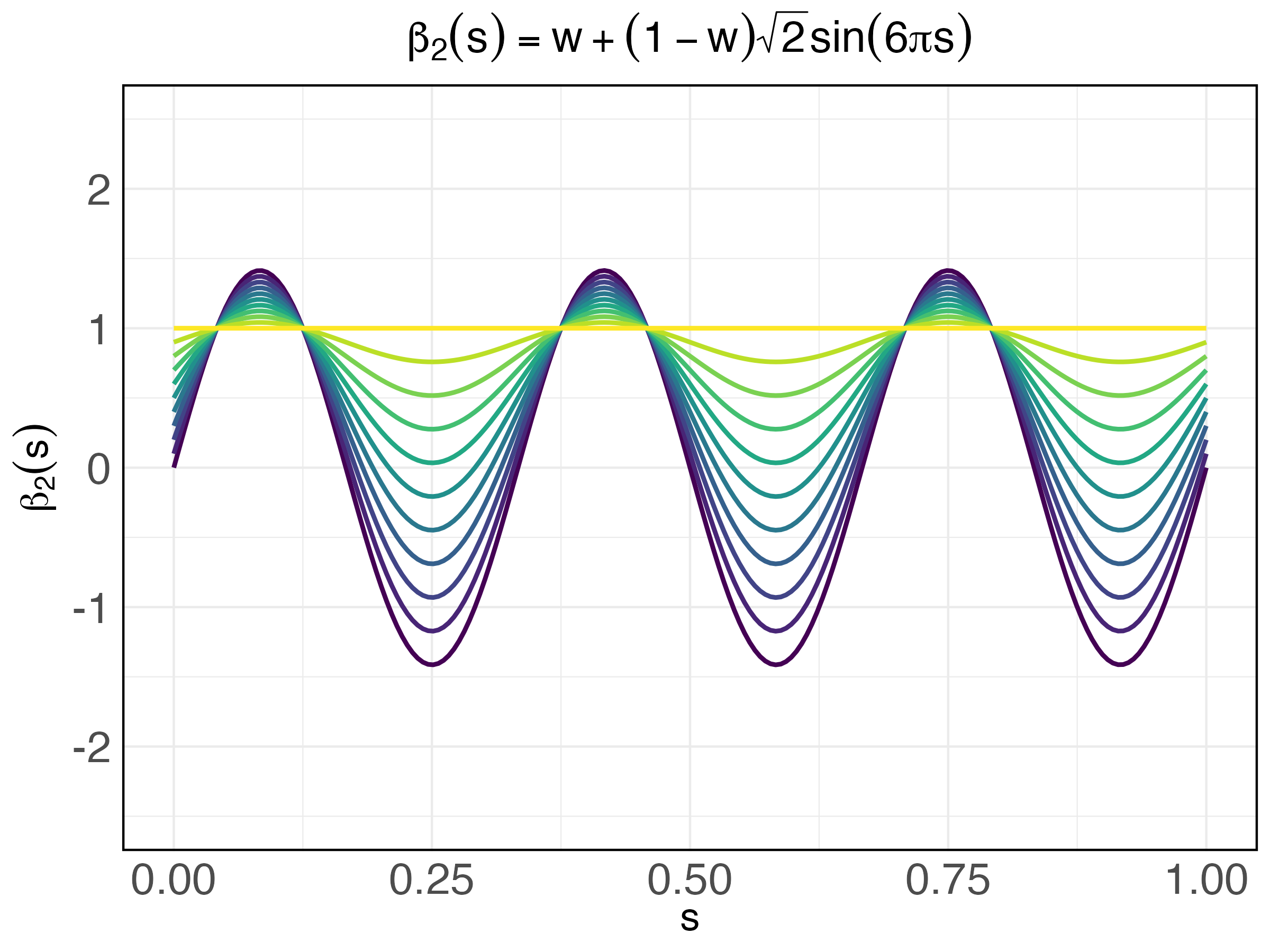}
& \includegraphics[width=4cm]{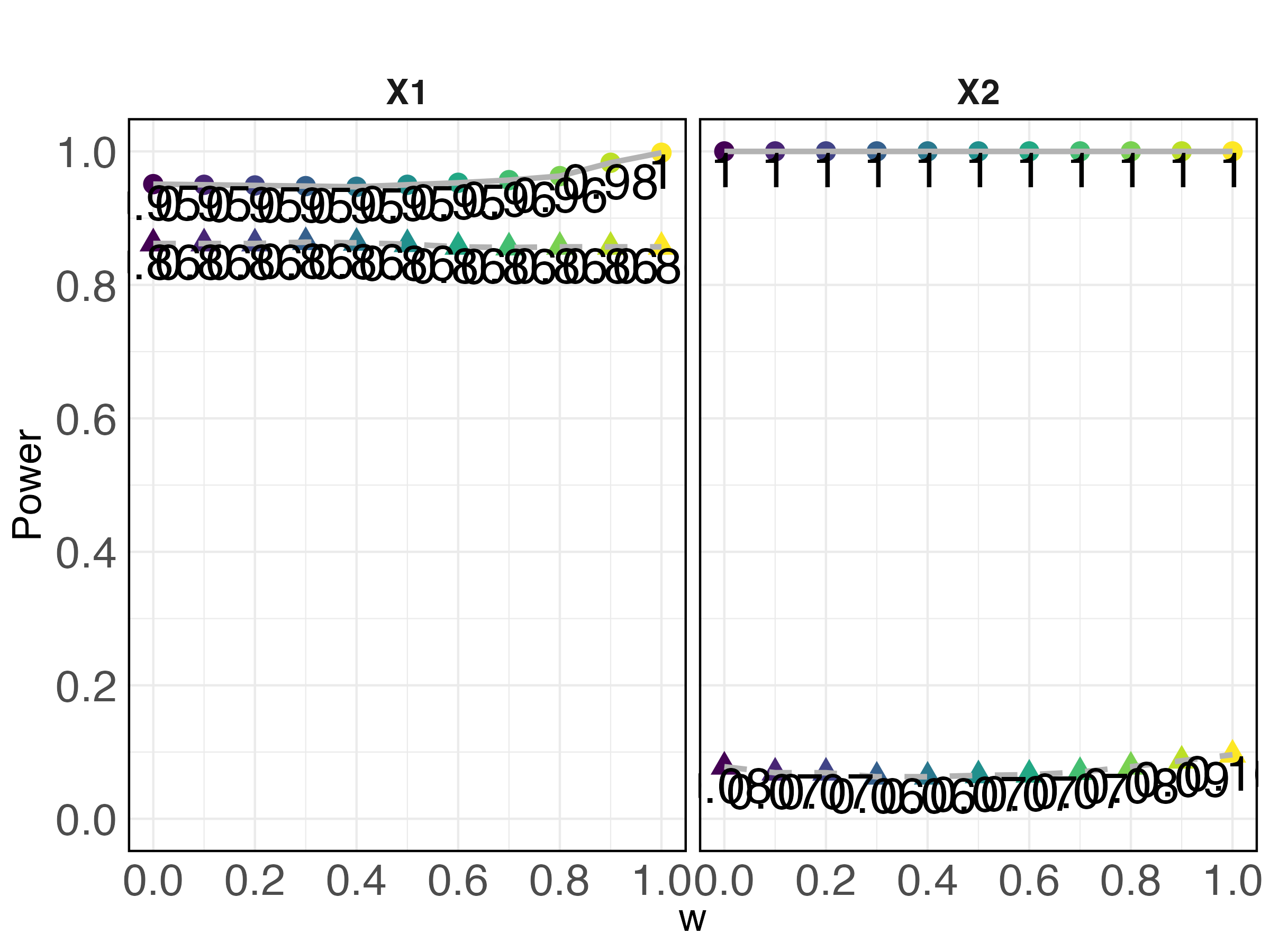} \\[2mm]
\hline
\end{tabular}
\caption{Simulation results for Scenario 2. For each cases, the second column illustrates $\beta_{1}(s)$ for different cases. The third column shows $\beta_{2,w}(s)$ as $w$ varies from $0$ to $1$. The fourth column shows the corresponding testing power curves for $\beta_1(s)$ and $\beta_{2,w}(s)$. Points are color-coded by $w$, with solid lines for the true-model test ($\mathbf{H}_0^{\text{TRUE}}$) and dashed lines for the misspecified-model test ($\mathbf{H}_0^M$).}
\label{fig:simu_power_2beta4phi_case1-6}
\end{figure}

Figure \ref{fig:simu_power_2beta4phi_case1-6} summarizes the empirical power for six cases (the labels in the first column indicate case number). Cases 1 through 3 correspond to $\beta_1(s) = 1$, while cases 4 through 6 correspond to $\beta_1(s)=\sqrt{2}\sin(2\pi s)$; the second column labeled $\beta_1(s)$ displays the corresponding functions of $\beta_1(s)$ for each case. Case 1 and 4 use $\beta_2(s) = w + (1 - w)\sqrt{2}\sin(2\pi s)$, cases 2 and 5 use $\beta_2(s) = w + (1 - w)\sqrt{2}\sin(4\pi s)$, and cases 3 and 6 use $\beta_2(s) = w + (1 - w)\sqrt{2}\sin(6\pi s)$. The third column labeled $\beta_2(x)$ displays the family of curves for $\beta_2(s)$, with colors varying from yellow for $w=0$ (perfect correlation with $\beta_1(s)$) to purple (zero correlation with $\beta_1(s)$). The fourth column labeled Power of $\beta_1(s)$, $\beta_2(s)$ displays the power of detecting $\beta_1(s)$ and $\beta_2(s)$ with different shape of $\beta_2(s)$.

For cases 1 through 3, FoSR has power close to $1$ for detecting $\beta_1(s)$ irrespective of the shape of $\beta_{2,w}(s)$, whereas the power of RPCS is close to the size of the test (no power). 
FoSR also has close to maximum power for detecting $\beta_{2,w}(s)$ for all cases. The power of RPCS tends to be higher for cases 1 and 2 when $w=0$ and $\beta_{2,w}(s)$ is collinear with the first and second principal component, respectively. Even in this best-case scenario, the power is smaller than that of FoSR, which could be due to the extra fixed effects parameter, $\beta_1(s)$, that is not accounted for by RPCS. As $w$ increases $\beta_{2,w}(s)$ becomes closer and closer to $\beta_1(s)$, which is orthonormal to the PCs, and power decreases rapidly to the size of the test (no power). In Case 3, the power of the RPCS is close to the size of the test for $\beta_{2,w}(s)$, irrespective of $w$. This is the case when $w$ induces a linear combination between two functions that are both orthonormal on all PCs.

For cases 4 through 6, the power to detect $\beta_1(s)$ is relatively constant for all values of $w$, with better power for FoSR. For $\beta_{2, w}(s)$, the power of RPCS is always lower than that of FoSR and it: (1) decreases fast to the size of the test (no power) as a function of $w$ in cases 4 and 5; (2) is equal to the size of the test for all $w$ for case 6. Recall that in cases 4 through 6, $w=1$ corresponds to $\beta_{2, w}(s)=1$, which is orthonormal to all four PCs. In case 4, $w=0$ corresponds to $\beta_{2, w}(s)=\sqrt{2}\sin(2\pi s)$, which is collinear with PC1; in case 5, $w=0$ corresponds to $\beta_{2, w}(s)=\sqrt{2}\sin(4\pi s)$, which is collinear with PC2; and in case 6, $w=0$ corresponds to $\beta_{2, w}(s)=\sqrt{2}\sin(6\pi s)$, which is orthonormal to all PCs. In all scenarios, RPCS has no power when the fixed effects are orthonormal to the PCs, and has much lower power than FoSR, even in ideal situations for RPCS. RPCS rapidly loses power as the correlation between the fixed effects and PCs is not $1$ (when the shape of the two functions is not identical.)  We conclude that our simulations exactly correspond to our theoretical results and discussion in Section~\ref{subsec:power_loss}.

‘\section{Application to the NHANES accelerometry study}\label{sec:real_data}
\begin{figure}[!tbh]
\centering
\includegraphics[width=\textwidth]{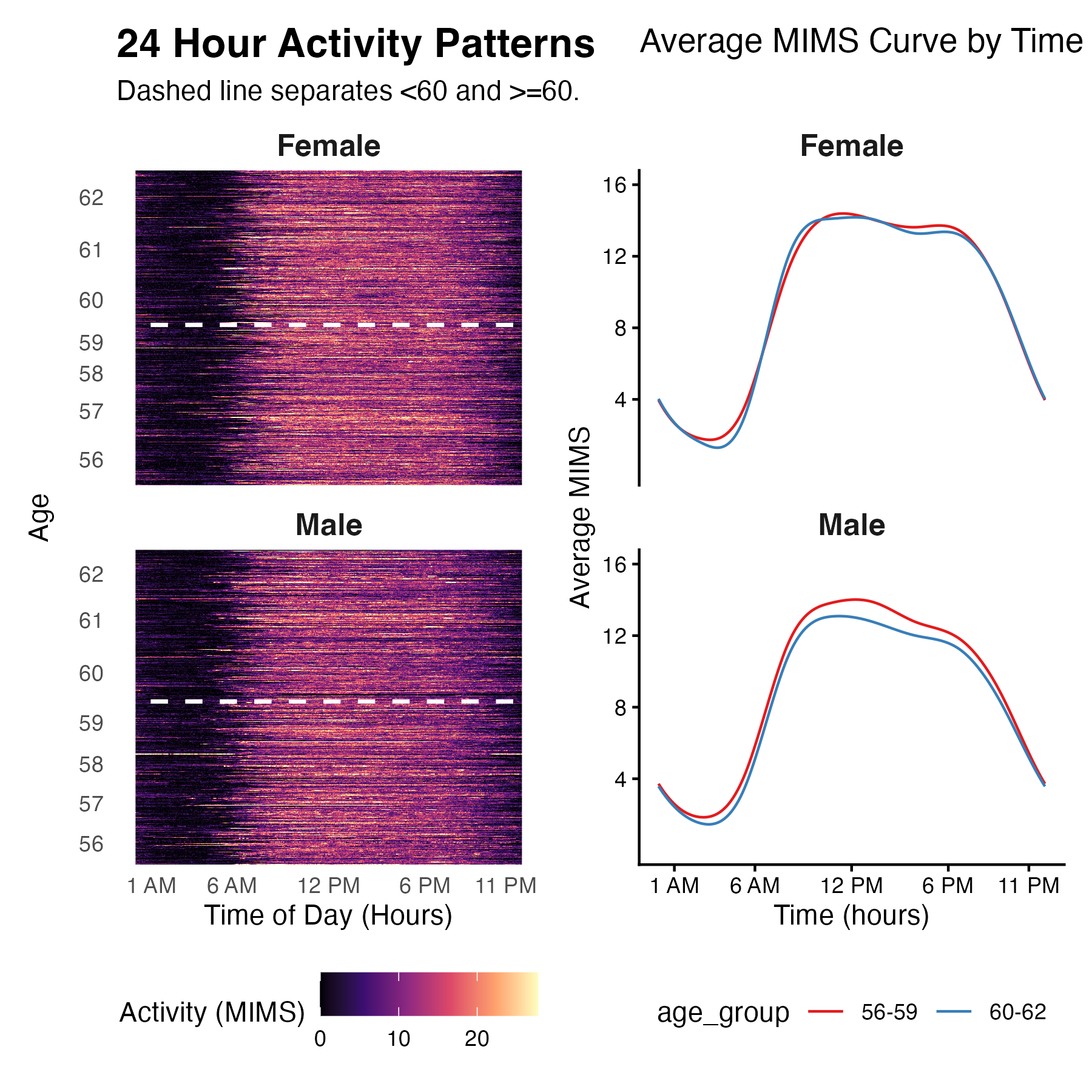}
  \caption{Left panels: individual physical activity (MIMS value) for all study participants aged 56-62, split by gender. Brighter colors imply more intense activity for a minute. The white dashed line separates subjects with ages $<60$ and $\geq 60$. Right panels: smoothed average physical activity (MIMS value), color-coded by the gender group, red for 56-59 and blue for 60-62.} 
  \label{fig:nhanes_data}
\end{figure}
To examine the potential for power loss in dense functional data, we utilize minute-level physical activity data obtained from wrist-accelerometers from the 2011–2014 National Health and Nutrition Examination Survey (NHANES). The data consist of Monitor Independent Movement Summary (MIMS) units \citep{john2019mims}, a validated measure of activity intensity. These data are densely sampled on a regular grid, with $1{,}440$ observations per participant—one for every minute of a $24$-hour cycle (midnight to midnight). We conduct first the analysis for participants aged $56$–$62$ ($n=925$) and the accelerometry data are displayed in Figure \ref{fig:nhanes_data}. The left panels display the Swihart-lasagna plots \citep{swihart2010} for the minute-level MIMS separated by sex; each row corresponds to a study participant ordered by age (higher corresponds to older study participants). The x-axis is time from midnight to midnight. A brighter color indicates higher MIMS (more intense physical activity) and the white dashed line separates study participants with ages $<60$ and $\geq 60$. The right panels display the age-specific group means (red corresponds to the $56$-$59$ age group and blue to the $60-62$ age group, respectively) separated by sex (females in the top-right panel and males in the bottom-right panel). For example, the blue line in the top-right panel is obtained by taking the column means of the data in the left-top panel above the dashed white line.  

The average daily physical activity for males (bottom-right panel) is visibly lower for older males ($60$-$62$ years of age), especially during typical wake hours ($6$AM to $11$PM). A similar effect may also be present among females, though the the differences are more subtle within this age range. Of course, these are estimators and it is unclear whether these observed results are unusual given the underlying variability of the data (hinted at in the left panels of Figure \ref{fig:nhanes_data}, but not shown in full because of the difficulties of plotting high dimensional functional data.)

\begin{figure}[ht]
\centering
\includegraphics[width=16cm]{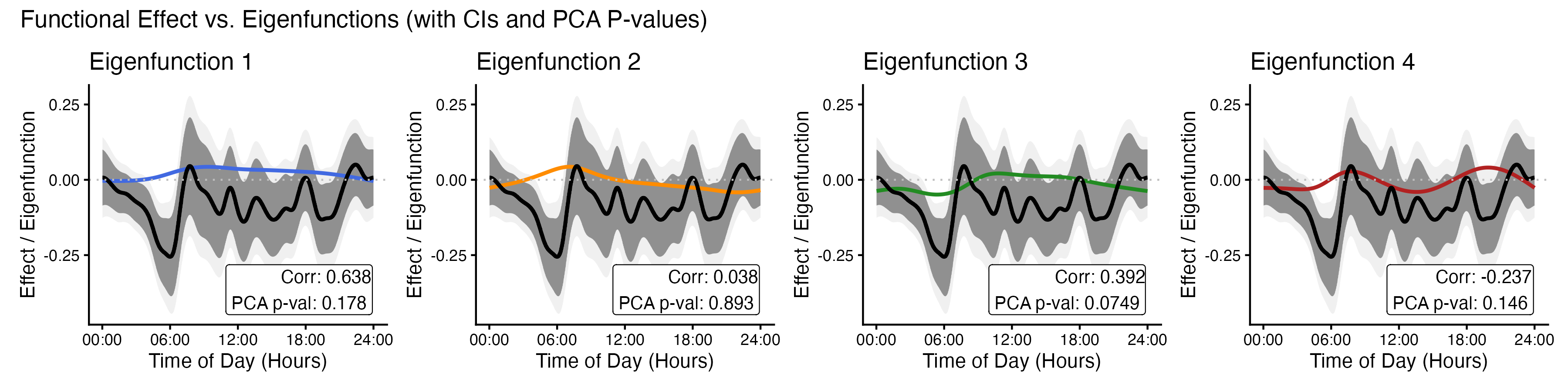}
\caption{The four panels display estimated eigenfunctions from FPCA, respectively (solid lines, blue for first PC, orange for second PC, green for third PC, and red for fourth PC) and estimated age effect from FoSR (black lines), together with their $95$\% pointwise (dark gray shaded areas) and correlation and multiplicity adjusted (CMA, light gray shaded areas) confidence intervals. The $p$-values for the $t$-test of the null hypotheses $H_{0,{\rm age}}^M:b_{{\rm age},l}=0$ and the correlations between this estimated effect and the estimated principal components are shown in the bottom right of the four panels.}
\label{fig:nhanes_age_effect_comparison}
\end{figure}

We apply and compare RPCS and FoSR to conduct inference on the age effect, adjusted for BMI and sex (results are similar without adjustment). We first performed FPCA on the physical activity data by fitting the misspecified model~\eqref{eq:misspecified}. The age effect is then tested by regressing these PC scores ($\xi^M_{il}$) on age, adjusting for gender and BMI: $\xi^M_{il} = b_{0,l} + b_{{\rm age},l} \text{Age}_i + b_{{\rm sex},l} \text{sex}_i + b_{{\rm BMI},l}\text{BMI}_i + e_{il}$ for $l = 1, \ldots, 4$. The first four estimated eigenfunctions explain $89\%$ of the variability, displayed in the four panels of Figure \ref{fig:nhanes_age_effect_comparison}, respectively (solid lines, blue for first PC, orange for second PC, green for third PC, and red for fourth PC). The amounts of variation explained are $51\%$ by PC1, $22\%$ by PC2, $10\%$ by PC3, and $6\%$ by PC4. The $p$-values for the $t$-test of the null hypotheses $H_{0,{\rm age}}^M:b_{{\rm age},l}=0$ are $0.178$ for PC1 ($l=1$), $0.893$ for PC2 ($l=2$), $0.0749$ for PC3 ($l=3$), and $0.146$ for PC4 ($l=4$), even without accounting for multiplicity (the p-values using Bonferonni corrections are larger). These p-values are shown in the bottom right corner of each panel of Figure~\ref{fig:nhanes_age_effect_comparison}. Thus, based on these results, one would conclude that there is no statistically significant effect of age on physical activity in this age range.

We compare results to those obtained using FoSR applied to the the model \eqref{eq:DGM} with three time-varying fixed effects (age, gender, and BMI) following the methods described in \citep{crainiceanu2024book}: $W_i(s) = \beta_0 + {\rm age}_{i}\beta_{\rm age}(s) + {\rm sex}_{i}\beta_{\rm sex}(s) + BMI_i \beta_{\rm BMI}(s) + \sum_{l = 1}^4 \xi_{il}\phi_l(s) + \varepsilon_{i}(s)$. The estimated functional age effects $\beta_{\rm age}(s)$ are displayed as the black curves in each panel of Figure \ref{fig:nhanes_age_effect_comparison} together with their $95$\% pointwise (dark gray shaded areas) and correlation and multiplicity adjusted (CMA, light gray shaded areas) confidence intervals \citep{crainiceanu2024book,Crainiceanu2012}. The CMA confidence intervals do not include zero during a long period in the morning (approximately from $4$ AM until $7$ AM), as well as for brief periods in late afternoon and early evening. The CMA-corrected p-values for this period are $<0.001$. Overall, the point estimators are negative, which corresponds to lower activity at higher ages, much lower (and statistically significant) activity very early in the morning. Therefore, FoSR detects strong effects when RPCS does not and provides translatable details that cannot be reproduced via RPCS.  

Our results indicated that the loss of power is likely affected by the correlations between the true effect and the principal components of the data without modeling the true functional effect. In this example, we do not have the true effect, but we have the estimated one, $\widehat{\beta}_{\rm age}(s)$. The correlations between this estimated effect and the estimated principal components are $0.638$ (with PC1), $0.038$ (with PC2), $0.392$ (with PC3), and $-0.237$ (with PC4), respectively; results are shown in the bottom right of the four panels of Figure~\ref{fig:nhanes_age_effect_comparison}. These correlations play a crucial role in reducing the power of the tests; note that it is currently unknown how to reconstruct $\widehat{\beta}_{\rm age}(s)$ or conduct inference from the PCA regression scores. Unfortunately, this is obvious in retrospective. 

\subsection{Sensitivity Analysis: Expanding the Age Group}\label{subsec:expansion}
\begin{figure}[ht]
\centering
\begin{tabular}{|c|c|c|c|}
\hline
Age Group & $b_{age,l}$ Projection \\
\hline
56-60 & \includegraphics[width=13.5cm]{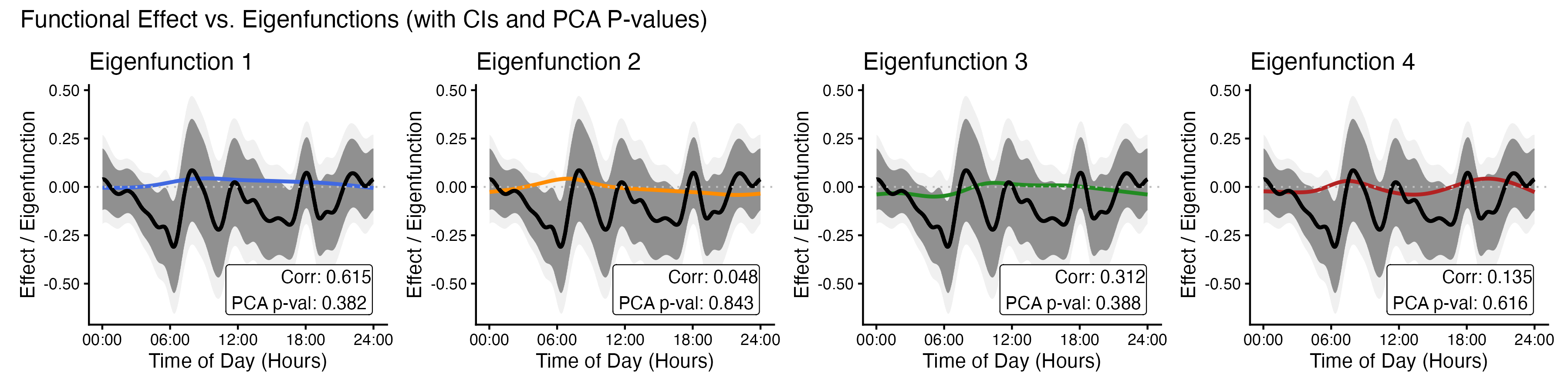} \\[2mm]

56-62 & \includegraphics[width=13.5cm]{figures/nhanes/correlation_55_62.png}\\[2mm]

56-65 & \includegraphics[width=13.5cm]{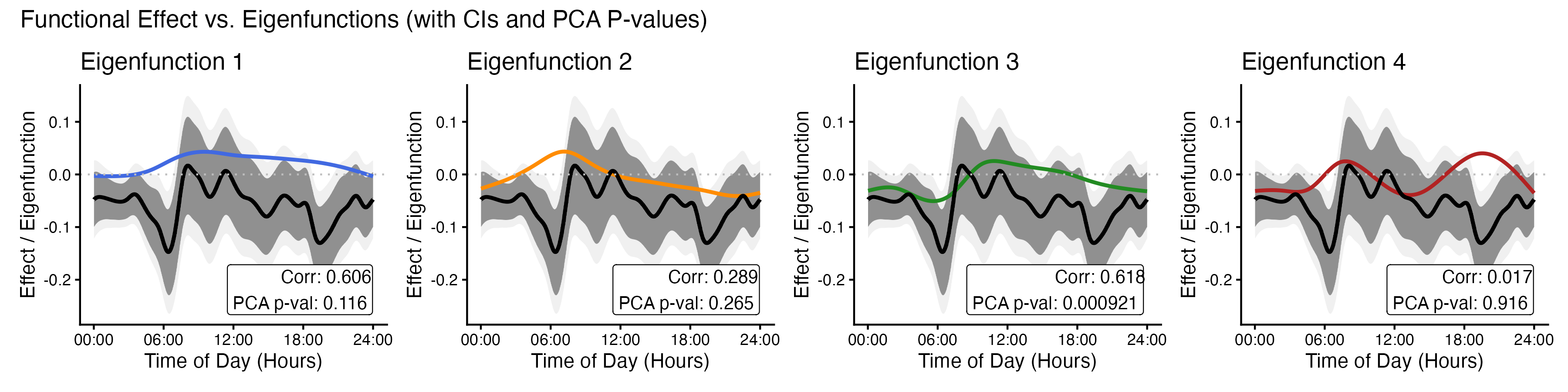}\\[2mm]
\hline
\end{tabular}
\caption{Estimated eigenfunctions from FPCA and estimated age effect from FoSR for expanding age groups. Middle column: The four panels display the first four eigenfunctions (solid lines: blue for PC1, orange for PC2, green for PC3, and red for PC4). The black line in each panel represents the estimated age effect from FoSR, accompanied by $95$\% pointwise (dark gray) and correlation/multiplicity adjusted (CMA, light gray) confidence intervals. The $p$-values for the $t$-test of the null hypotheses $H_{0,\text{age}}^M:b_{\text{age},l}=0$ and the correlations between this estimated effect and the estimated principal components are shown in the bottom right of the respective panels. Right column: Age effect estimated by FoSR, reconstructed from all 4 PCs and reconstructed using only the significant PCs.}
\label{fig:nhanes_age_group_5565}
\end{figure}

We conducted a sensitivity analysis across three age groups by slowly increasing the age range: (1) ages 56–60 ($n=635$), (2) ages 56–62 ($n=925$), and (3) ages 56–65 ($n=1,373$). Results are presented in Figure~\ref{fig:nhanes_age_group_5565}.
This comparison reveals a clear difference in the sensitivity of the two approaches. In the smallest group corresponding to strictest age restrictions (ages $56–60$), neither method identifies a significant age effect; the FoSR confidence intervals include zero throughout the 24-hour cycle, and all RPCS results yield large $p$-values. As the cohort expands to ages $56–62$, the RPCS still fails to detect an association across the first four principal components (PCs). Conversely, the FoSR model identifies a significant age effect between 4:00 AM and 6:30 AM, indicating that older individuals exhibit lower early-morning activity levels. These results were shown earlier, and rediscussed here for comparison purposes across age groups. In the largest group corresponding to ages $56–65$, the PCA approach detects a signal in the PC3 (p-value $<0.001$). Simultaneously, the FoSR model identifies more pronounced effects, revealing significant activity differences in both the early morning and evening (6:30 PM to 9:00 PM). These findings demonstrate that RPCS requires a substantially larger signal to detect an effect, whereas FoSR more reliably isolates associations by directly modeling the functional coefficient structure. Moreover, even in this case when RPCS detects a signal, it remains largely unknown where the differences are during the day and what are the practical implications of such a finding. Such are the toils of RPCS: easy to do and hard to interpret.

\subsection{Sensitivity Analysis: Non-overlapping Age Groups}\label{subsec:expansion}
\begin{figure}[!tbh]
\centering
\includegraphics[width=\textwidth]{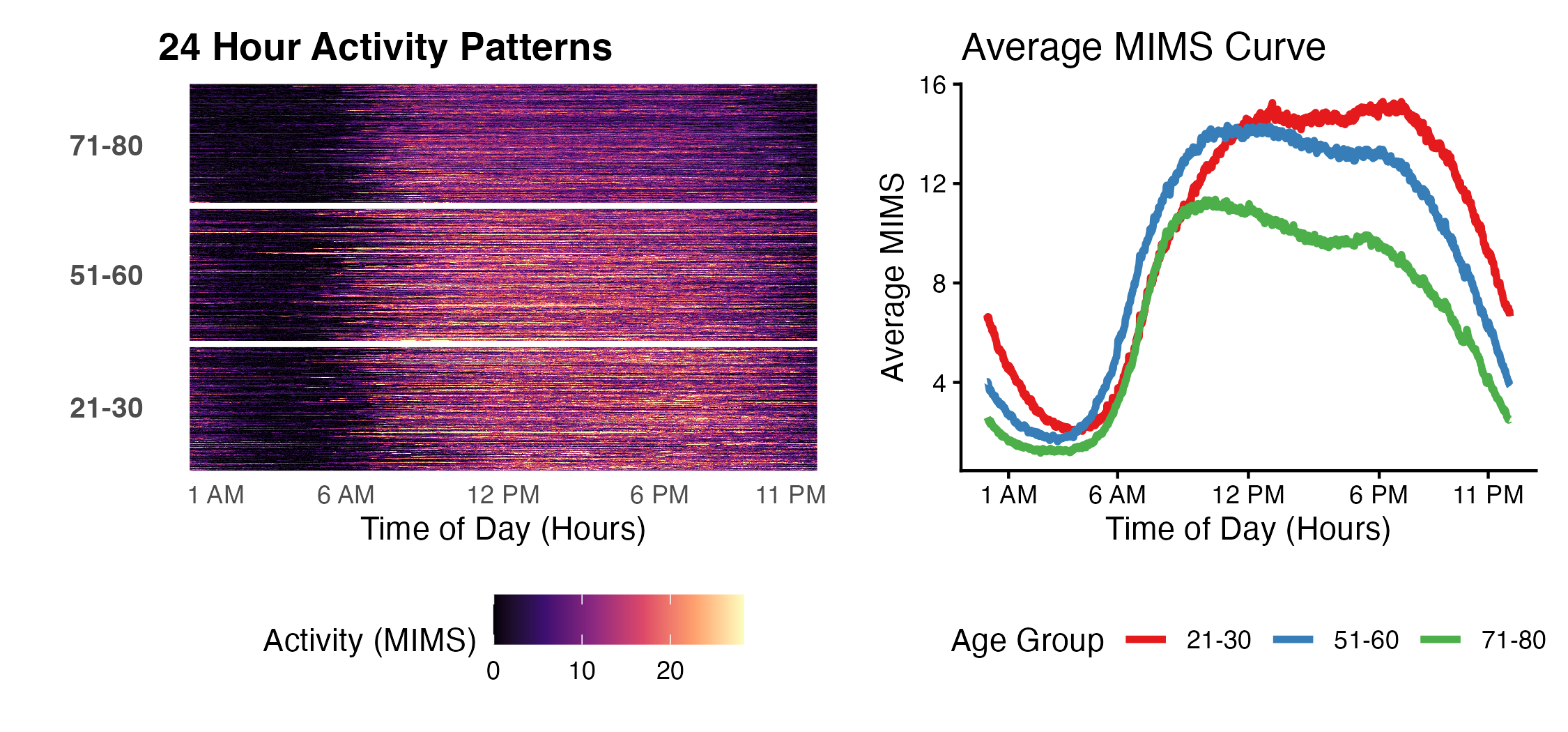}
  \caption{Left panels: individual physical activity (MIMS value) for all study participants aged 21-30, 51-60 and 71-80, separated by age groups. Brighter colors correspond to higher intensity of physcal activity. Right panels: average physical activity (MIMS value), color-coded by  age group, red for 21-30, blue for 51-60 and green for 71-80.} 
  \label{fig:nhanes_data_age_group}
\end{figure}

To further assess the detection of age effects across distinct life stages, we analyzed three non-overlapping 10-year age intervals: ages 21–30 ($n=1,183$), 51–60 ($n=1,290$), and 71–80 ($n=1,138$). The 24-hour activity patterns for these groups are illustrated in Figure~\ref{fig:nhanes_data_age_group}. The left panels display individual physical activity (MIMS values), showing that while all subjects are predominantly active during typical daytime hours (6:00 AM to 11:30 PM), notable differences can be observed across these age subgroups. The 71–80 subgroup exhibits the lowest overall activity (darker color) and the earliest sleep onset. Within this older group, increasing age is associated with progressively lower activity and earlier sleep times. The 51–60 subgroup displays a more consistent activity pattern, while the 21–30 cohort exhibits the highest nighttime activity (see the higher average physical activity between 12AM and 3AM). Minute-level averages (Figure~\ref{fig:nhanes_data_age_group}, right panels) further clarify these trends: the 21–30 cohort is most active from the afternoon into the late night, the 51–60 cohort peaks in the morning (4:30 AM to 12:00 PM), and the 71–80 cohort remains the least active throughout the entire day. 

\begin{figure}[htbp]
    \centering
    \resizebox{\textwidth}{!}{
    \begin{tabular}{|c|c|c|}
    \hline
    Age Group & $b_{age,l}$ Projection & Age Effect Estimation\\
    \hline
    21-30 & \includegraphics[width=0.7\textwidth]{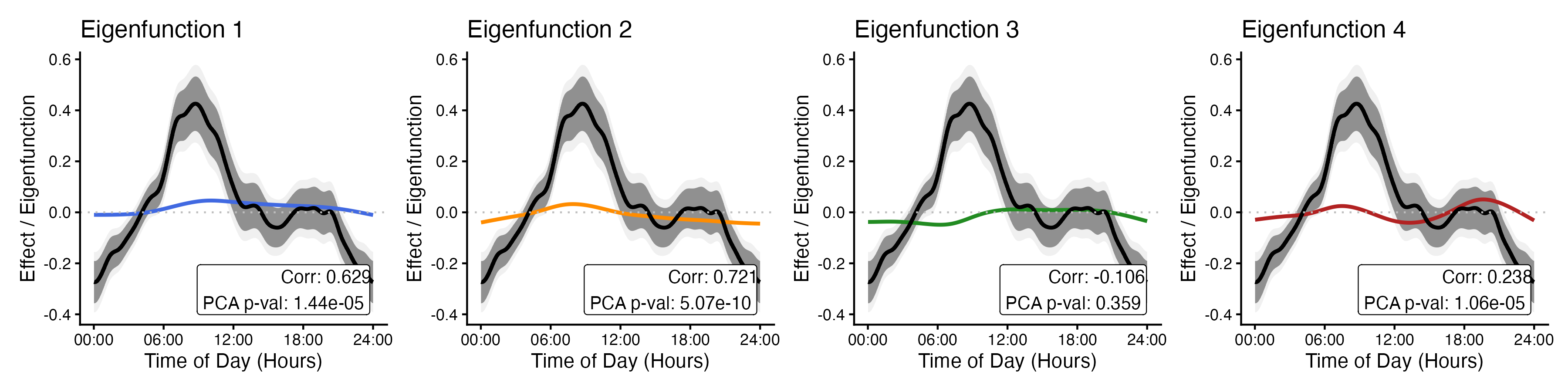} & \includegraphics[width=0.2\textwidth]{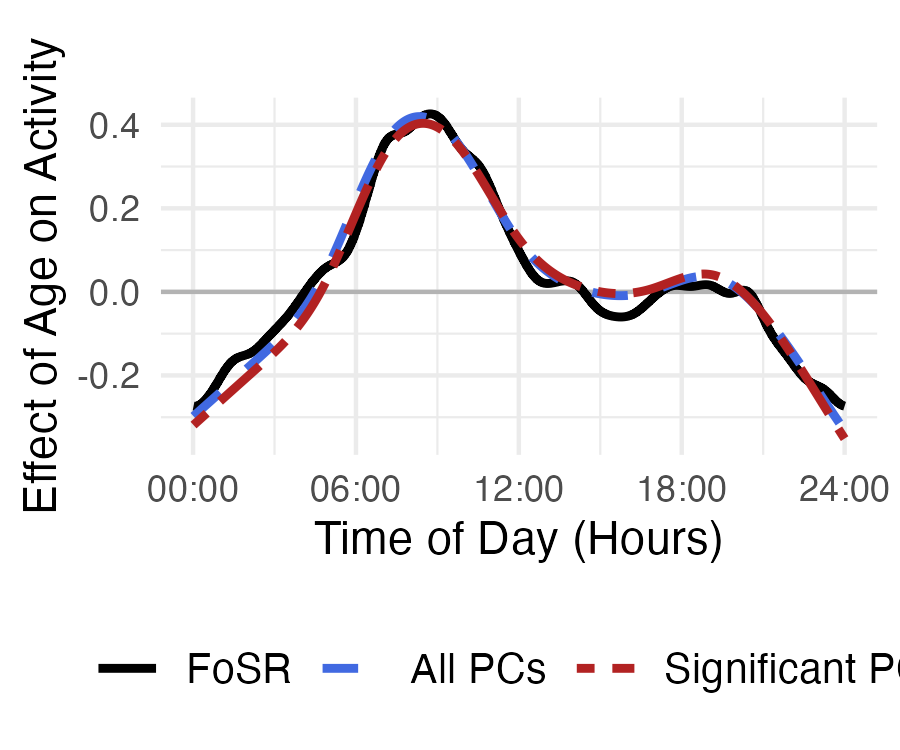} \\[2mm]
    
    51-60 & \includegraphics[width=0.7\textwidth]{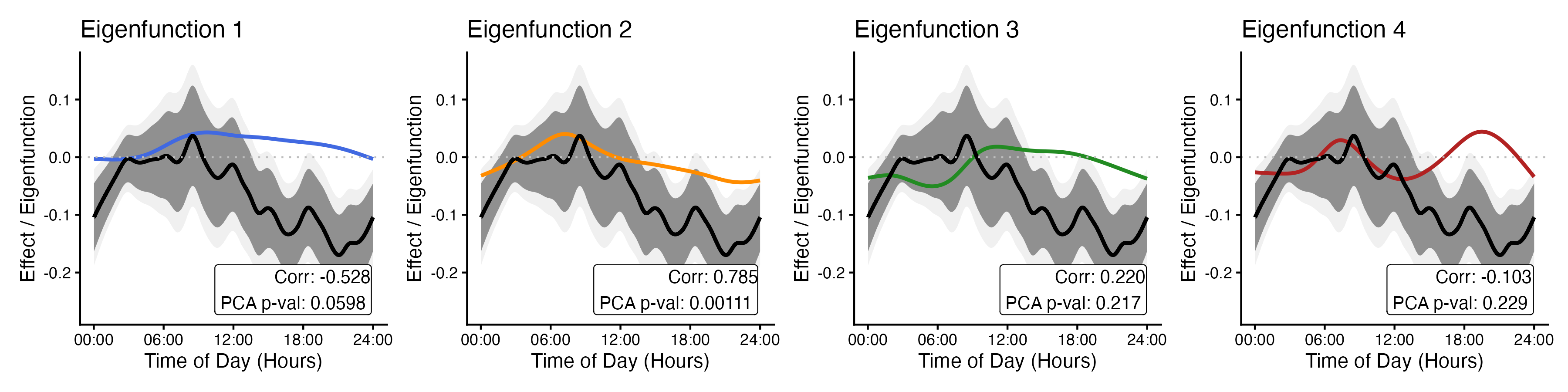}& \includegraphics[width=0.2\textwidth]{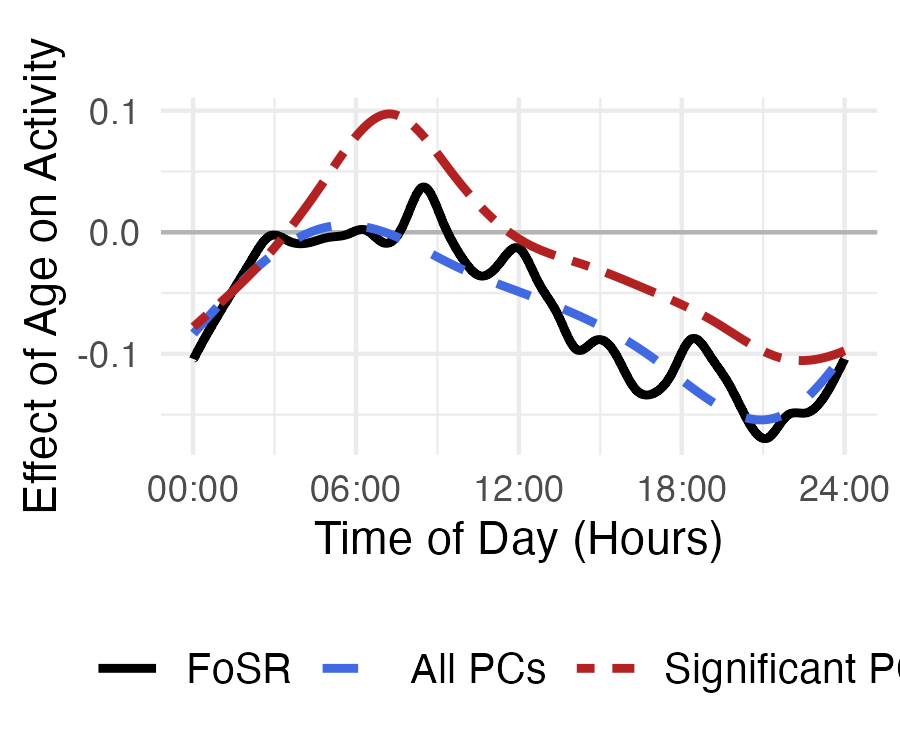}\\[2mm]
    
    71-80 & \includegraphics[width=0.7\textwidth]{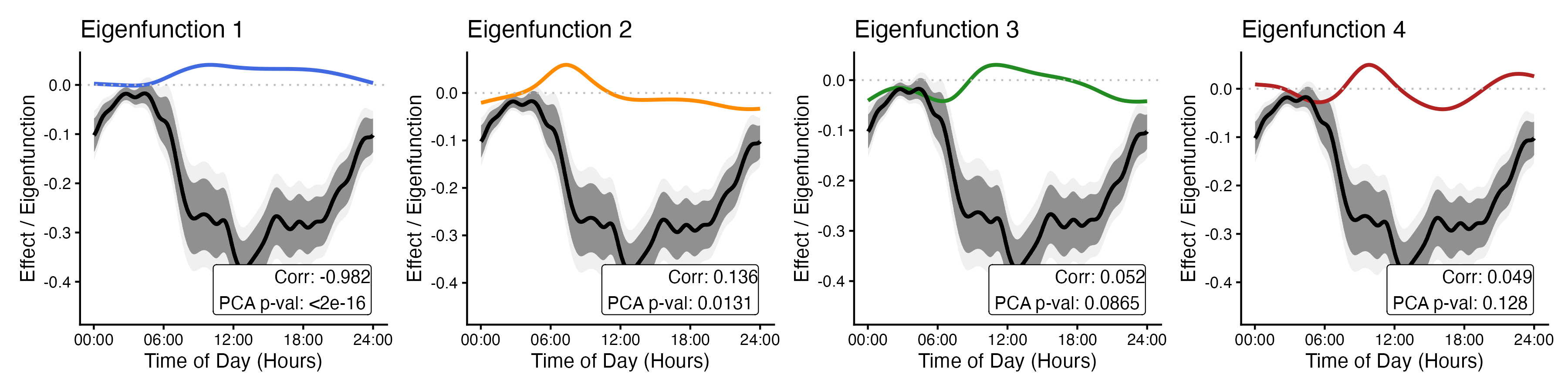}& \includegraphics[width=0.2\textwidth]{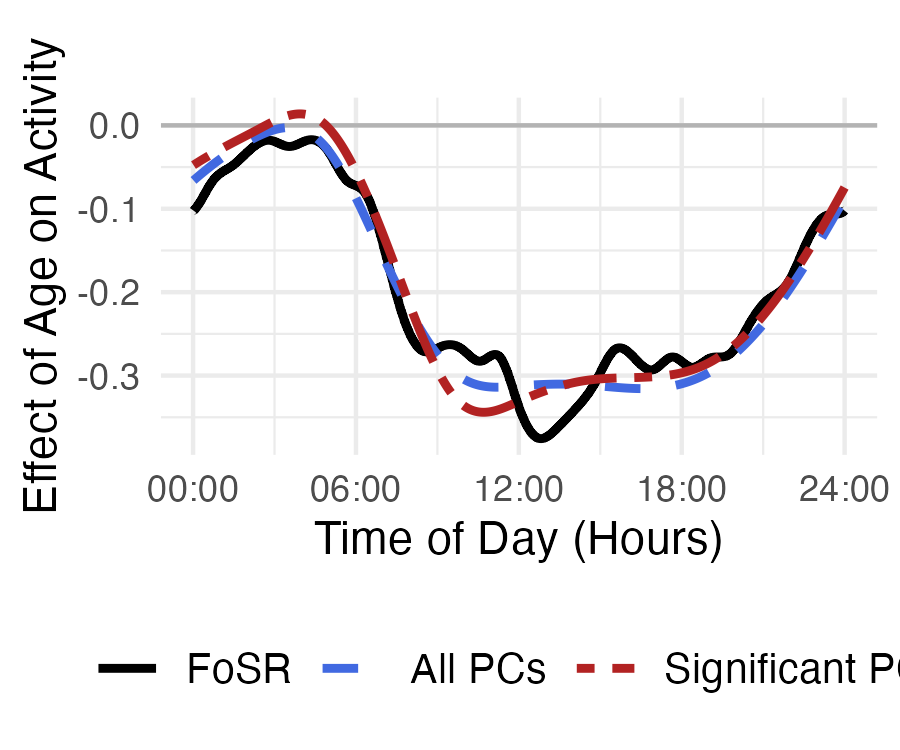}\\[2mm]
    \hline
    \end{tabular}
    }
    \caption{Estimated eigenfunctions from FPCA and estimated age effect from FoSR for non-overlapping age groups. Middle column: The four panels display the first four eigenfunctions (solid lines: blue for PC1, orange for PC2, green for PC3, and red for PC4). The black line in each panel represents the estimated age effect from FoSR, accompanied by $95$\% pointwise (dark gray) and correlation/multiplicity adjusted (CMA, light gray) confidence intervals. The $p$-values for the $t$-test of the null hypotheses $H_{0,\text{age}}^M:b_{\text{age},l}=0$ and the correlations between this estimated effect and the estimated principal components are shown in the bottom right of the respective panels. Right column: Age effect estimated by FoSR, reconstructed from all 4 PCs and reconstructed using only the significant PCs.}
    \label{fig:nhanes_age_group_seq}
\end{figure}

We applied both RPCS and FoSR to each age group separately, with results presented in Figure~\ref{fig:nhanes_age_group_seq}. While the empirical activity profiles differ substantially across the three groups, their first four estimated eigenfunctions are quite consistent across age groups. This suggests that the fundamental "building blocks" of daily activity variation are universal, regardless of age. In contrast, the FoSR estimates of the age effects exhibit strong variation across age groups, indicating that the impact of getting older on daily patterns of activity is highly specific to each life stage. For the 21–30 cohort, the score coefficients for PC1 ($p = 1.44 \times 10^{-5}$), PC2 ($p = 5.07 \times 10^{-10}$), and PC4 ($p = 1.06 \times 10^{-5}$) indicate strong evidence against the null hypothesis of no age effect. Their correlations with the FoSR estimate $\widehat{\beta}_{\text{age}}(s)$ are relative large ($0.629$, $0.721$, and $-0.238$, respectively). This shows that the overall age-related changes in this young adult group are driven by complex changes in multiple physical activity patterns. This is likely reflective of the major, multifaceted lifestyle transitions that occur during this decade. 
For the 51–60 cohort, only the association between the score on the PC2 and age is significant ($p = 0.001$), which is likely due to the larger correlation between PC2 and the FoSR estimated age effect (correlation = $0.785$). A subject with a positive score in PC2 is more active in the morning and less active in the afternoon and at night. The FoSR estimated age effect is showing an older subject in this age group is less active in the afternoon to night time (3 PM to 1 AM). While PC1 represents mostly the change of total activity throughout the day, the significant association between the scores on the PC2 and age suggests that during middle age, overall activity levels remain relatively stable but the primary effect of aging is subtler and happens primarily during the afternoon and night time.
For the 71–80 subgroup, PC1 is very strongly correlated with the FoSR estimate of the age effect (correlation = $-0.98$) and the the association between scores on PC1 and age is strongly significant ($p < 2 \times 10^{-16}$). Because the first principal component captures the overall volume of physical activity (a subject with a positive score is more active throughout the day, and much more active during the afternoon to evening), this near-perfect negative correlation reveals a biological reality of this life stage: for individuals in their seventies, the age effect is almost entirely defined by a generalized, uniform decline in overall physical activity, with a stronger decline during the afternoon. This specific age effect is captured by FoSR, which is highly correlated with PC1.

We also illustrate the difficulty of reconstructing the age effects from RPCS. To do that we consider two plausible reconstructions of the functional effect of age: (1) using all four eigenfunctions irrespective to whether or not their scores were significantly associated with age (at an $\alpha=0.05$ level): $\tilde{\beta}_{\text{age}}(s) = \sum_{l=1}^4 \widehat{b}_{\text{age},l} \widehat\phi_l(t)$; and (2) restricted to only the eigenfunctions whose scores were significantly associated with age (at the same $\alpha=0.05$ level): $\tilde{\beta}^{'}_{\text{age}}(s) = \sum_{l: p_{\text{age},l}<0.05} \widehat{b}_{\text{age},l} \widehat\phi_l(t)$. The last panel column in Figure~\ref{fig:nhanes_age_group_seq} displays the FoSR estimator (black solid line), the estimator reconstructed from all four PCs (blue dashed line), and the estimator reconstructed from the PCs identified by regression (red, dashed and dotted line) for the $21-30$ (top panel), $51-60$ (middle panel), and $71-80$ (bottom panel) age groups, respectively. All point estimators for the $21-30$ and $71-80$ look similar, though some differences are apparent in the $71-80$ age groups. In this group, the PC-based estimators tend to be smoother, likely because of the smoothing induced by choosing a particular number of eigenfunctions, which happen to be the smoothest ones. In the $51-60$ age group the estimators based on all PCs and on only significant PCs are substantially different. In the absence of the FoSR estimator, there would be no objective way of choosing one versus another estimator. Having multiple estimators based on different strategies, provides a practical approach to estimation, as one substantially different estimator (such as the red line for the $51-60$ age range) may point to incorrect estimation of the true effect. Even if the point estimators sometimes agree between these methods, with the exception of FoSR, we are not aware of any methods for conducting inference on the reconstructed estimators from PC regressions.

\section{Discussion}\label{sec:discussion}
In this paper, we formalized the statistical properties of the widely used two-step RPCS approach and compared it with the correctly specified Function on Scalar Regression (FoSR). We demonstrated that RPCS is statistically equivalent to fitting a misspecified model. The primary consequence of this misspecification is a severe and unrecognized loss of statistical power to detect the associations between scalar predictors and functional outcomes.

The key of this issue, as established in Theorem~\ref{thm:thm1}, lies in the fundamental misalignment of modeling objectives. Principal component analysis identifies directions of maximal marginal variance in the functional response, irrespective of the covariates. Consequently, the estimand in PCA score regression is merely the projection of the true effect onto the estimated eigenvectors. If the true effect is not perfectly correlated to one of the leading principal components, RPCS loses power. Furthermore, relying only on statistically significant PCs for reconstruction often yields an estimated effect that drastically diverges from the true association. Testing multiple score coefficients $b_{ql}$ also introduces testing multiplicity problems, making the decision of how many PCs to retain practically difficult and the interpretation of the fixed effect in the original functional domain highly problematic.

Our simulations and real-data applications underscore the practical dangers of this methodological blind spot. In the NHANES application, the RPCS approach completely masked the effect of age on early morning activity in the 56–62 age subgroup. These findings suggest that false negative and highly distorted effect shapes are likely pervasive in applications using FPCA score regression, particularly when the true signal is small to moderate or is not perfectly correlated with one of the leading PCs of the functional data.

The implications of this work extends beyond far beyond functional data analysis and illustrates common, largely ignored problem in applications. Indeed, note that the word ``functional'' throughout the paper was only used because it makes sense in our application. The exact same modeling framework, and consequently the exact same vulnerability to power loss, is present in any application where the outcome is high dimensional and the predictors are scalar (e.g., genomics or imaging). Historically, the two-step PCA approach was favored for its computational simplicity and its reliance on familiar scalar linear regression. However, our findings demonstrate that this convenience comes at a cost to statistical power, signal reconstruction, and signal inference.

\section{Appendix}\label{sec:Appendix}

\subsection{Proof of Theorem~\ref{thm:thm1}}

Assume the covariates $\{X_{iq}\}$ are mutually independent and independent of the true random effects $\{\xi_{il}\}$ and measurement errors $\{\epsilon_i(s)\}$,  $\xi_{il} \sim N(0, \lambda_l)$ for $l=1,\ldots,L_{true}$, and $\epsilon_{i}(s)\sim N(0, \sigma_\epsilon^2)$, for $s\in S$. The misspecified model uses $L_{miss}$ principal components.

The scores $\xi_{il}^M$ from the misspecified model are:
$$
\xi_{il}^M = \int_S \left(W_i(s) - \mathbb{E}[W_i(s)] \right)\phi_l^M(s)ds.
$$

The conditional expectation of $\xi_{il}^M$ given $\mathbf{X}_i$ is
$$\mathbb{E}[\xi_{il}^M \mid \mathbf{X}_i] = \mathbb{E}[\int_S \left(W_i(s) - \mathbb{E}[W_i(s)] \right)\phi_l^M(s)ds\mid \mathbf{X}_i]$$

Since the expectation is linear and the covariates are independent of the random effects and errors, $\mathbb{E}[\xi_{il}^M \mid \mathbf{X}_i]$ can be simplified to 
\begin{align*}
    \mathbb{E}[\xi_{il}^M \mid \mathbf{X}_i] &= \mathbb{E}[\int_S \left(W_i(s) - \mathbb{E}[W_i(s)] \right)\phi_l^M(s)ds\mid \mathbf{X}_i] \\
    &= \int_S \left(\mathbb{E}[W_i(s) \mid \mathbf{X}_i] - \mathbb{E}[\mathbb{E}[W_i(s)] \mid \mathbf{X}_i] \right)\phi_l^M(s)ds \\
    &= \int_S \left(\mathbb{E}[W_i(s) \mid \mathbf{X}_i] - \mathbb{E}[W_i(s)] \right)\phi_l^M(s)ds \\
    &= \int_S \left(\beta_0(s) + \sum_{q=1}^Q X_{iq}\beta_q(s) - \left(\beta_0(s) + \sum_{q=1}^Q \mathbb{E}[X_{iq}]\beta_q(s)\right)\right)\phi_l^M(s)ds \\
    &= \int_S \left(\sum_{q=1}^Q \left(X_{iq} - \mathbb{E}[X_{iq}]\right)\beta_q(s)\right)\phi_l^M(s)ds \\
\end{align*}

Now, consider the linear model for the scores, $\mathbb{E}[\xi_{il}^M \mid \mathbf{X}_i] = b_{0l} + \sum_{q=1}^Q b_{ql}X_{iq}.$
\begin{align*}
    b_{ql} &= \frac{\partial}{\partial X_{iq}}\mathbb{E}[\xi_{il}^M \mid \mathbf{X}_i] \\
    &= \frac{\partial}{\partial X_{iq}}\int_S \left(\sum_{q=1}^Q \left(X_{iq} - \mathbb{E}[X_{iq}]\right)\beta_q(s)\right)\phi_l^M(s)ds \\
    &= \int_S \frac{\partial}{\partial X_{iq}} \left(\sum_{q=1}^Q \left(X_{iq} - \mathbb{E}[X_{iq}]\right)\beta_q(s)\right)\phi_l^M(s)ds \\
    &= \int_S \beta_q(s)\phi_l^M(s)ds
\end{align*}

Testing $\mathbf{H}_0^M: b_{ql} = 0 \ \forall l$ is equivalent to testing whether $\beta_q(s)$ is orthogonal to all $\phi_l^M(s)$. This may fail to detect true signals ($\beta_q(s) \neq 0$) if $\beta_q(s)$ lies outside the span of $\{\phi_l^M(s)\}_{l=1}^{L_{miss}}$, leading to power loss.

\bibliographystyle{biom}
\bibliography{ref}

\end{document}